\documentclass[%
 reprint, 
 amsmath,amssymb,
 aps,
citeautoscript,
prb,
floatfix]{revtex4-1}
\usepackage[colorlinks=false]{hyperref}
\usepackage{float}
\usepackage{graphicx}%
\usepackage{dcolumn}%
\usepackage{bm}%
\usepackage{xcolor}
\usepackage[normalem]{ulem}
\usepackage{color,soul}
\usepackage{float} 
\usepackage[]{lineno}
\begin{document}
% \linenumbers

\preprint{APS/123-QED}

\title{Integrated frequency-modulated optical parametric oscillator}

% TBD
\author{Hubert~S.~Stokowski\textsuperscript{1}}
\author{Devin~J.~Dean\textsuperscript{1}}
\author{Alexander~Y.~Hwang\textsuperscript{1}}
\author{Taewon~Park\textsuperscript{1,2}}
\author{Oguz~Tolga~Celik\textsuperscript{1,2}}
\author{Marc~Jankowski\textsuperscript{1,3}}
\author{Carsten~Langrock\textsuperscript{1}}
\author{Vahid~Ansari\textsuperscript{1}}
\author{Martin~M.~Fejer\textsuperscript{1}}
\author{Amir~H.~Safavi-Naeini\textsuperscript{1,{$\star$}}\vspace*{3 mm}}

\affiliation{%
\textsuperscript{1} \mbox{Department of Applied Physics and Ginzton Laboratory, Stanford University, Stanford, California 94305, USA}
}%
\affiliation{%
\textsuperscript{2} \mbox{Department of Electrical Engineering, Stanford University, Stanford, California 94305, USA}
}%
\affiliation{%
\textsuperscript{3} \mbox{Physics \& Informatics Laboratories, NTT Research, Inc., Sunnyvale, California 94085, USA}
}%
\affiliation{%
\textsuperscript{{$\star$} } safavi@stanford.edu \vspace*{-6 mm}
}%

% \date{\today}

\pacs{Valid PACS appear here}
\maketitle

% Main text figures

\begin{figure*}[ht!]
  \begin{center}
      \includegraphics[width=\textwidth]{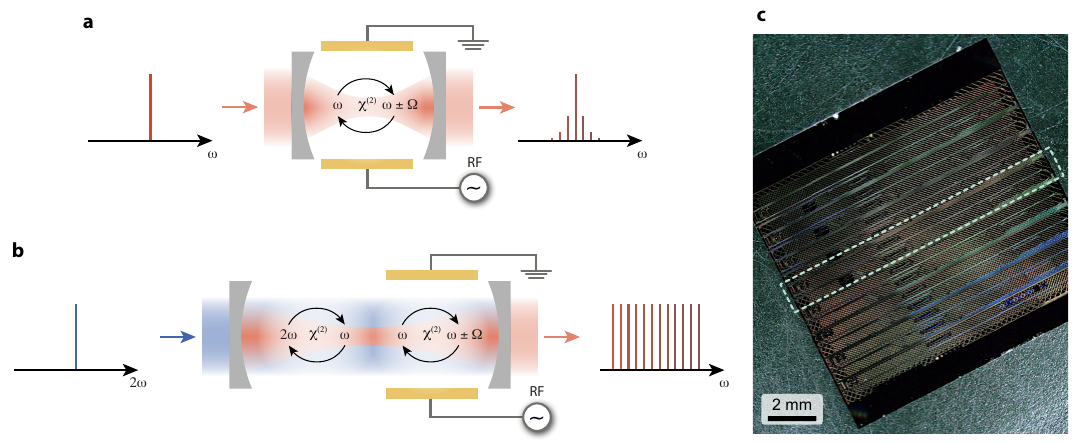}
  \end{center}
 \caption{\textbf{Frequency combs in the electro-optic resonator and frequency-modulated optical parametric oscillator. }
\textbf{a},~In a standard electro-optic comb, an optical resonator conducts electro-optic modulation on a single laser pump. This results in an output of equally spaced lines with intensities that decay exponentially.
\textbf{b},~The frequency-modulated optical parametric oscillator (FM-OPO) uses the second harmonic as a pump and emits light at the fundamental frequency. An electro-optic modulator couples the neighboring cavity modes, creating an evenly distributed output comb.
\textbf{c},~The microscope image shows a thin-film lithium niobate chip housing eight FM-OPO devices. One device has a footprint around 1x10~mm$^2$ (highlighted with a dashed rectangle).
}
 \label{fig:fig1}
\end{figure*}

\begin{figure*}[ht!]
  \begin{center}
      \includegraphics[width=\textwidth]{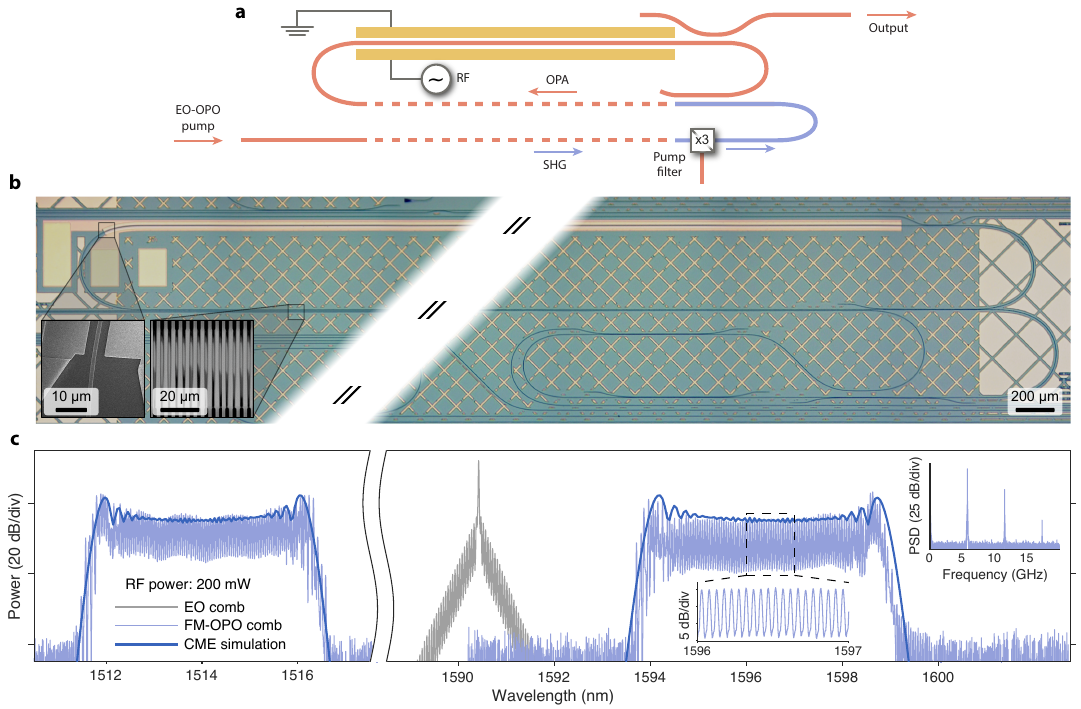}
  \end{center}
 \caption{\textbf{Frequency-modulated optical parametric oscillator (FM-OPO). }
\textbf{a},~The device features a racetrack resonator with an intracavity waveguide coupler, which allows only the fundamental harmonic to resonate while enabling the second harmonic to traverse the device once. One racetrack's straight section contains a periodically poled waveguide for second-order optical nonlinearity, while the other couples to microwaves through on-chip electrodes. A single-pass waveguide on the same chip generates the OPO's pump at the SH frequency, which is separated from the input C-band laser light using a series of filters.
\textbf{b},~The microscope image shows the FM-OPO racetrack resonator, with the center removed for clarity. The dark lines represent integrated waveguides. The yellow structure atop the racetrack is a gold micro-electrode slot waveguide that links with microwaves. The SEM micrograph in the left inset depicts the waveguide between the electrodes. The racetrack's bottom is periodically poled, as shown in the second harmonic microscope image of the poled film in the second inset. The waveguide circuit in the bottom right corner is a pump filtering section post the SHG waveguide.
\textbf{c},~The FM-OPO's output spectrum produces approximately 200 unique oscillator modes at the signal and idler frequencies under about 140 mW of input optical power and 200 mW of RF power. The dark blue line represents a coupled-mode theory simulation result. The bottom-right inset zooms into the comb's flat area, while the top-right inset displays an RF power spectral density of the FM-OPO output, detected using a fast photodiode with peaks at multiples of the cavity free spectral range.
}
 \label{fig:fig2}
\end{figure*}

\begin{figure*}[ht!]
  \begin{center}
      \includegraphics[width=\textwidth]{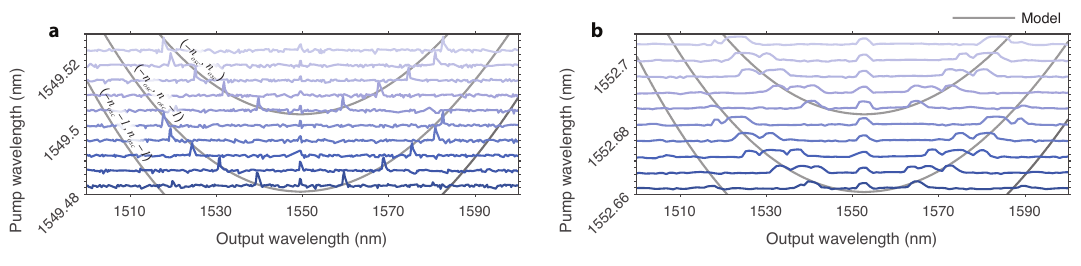}
  \end{center}
 \caption{\textbf{Pump-wavelength tuning of the FM-OPO. }
\textbf{a},~We examine the pump-wavelength tuning of a doubly resonant optical parametric oscillator. The blue traces relate to optical spectrum analyzer measurements. The DRO's output wavelength is determined by the frequency matching between the signal and idler modes with respect to the pump. The thin gray lines, recurring every 1/2 free spectral range, correspond to the theoretical model.
\textbf{b},~The pump-wavelength tuning of the FM-OPO aligns with the pure OPO trend. However, the entire frequency-modulated clusters of modes tune synchronously along the energy-conservation lines. For both \textbf{a} and \textbf{b}, one division on the y-axis corresponds to a 50 dB change in the OSA measurements.
}
 \label{fig:fig3}
 
\end{figure*}

\begin{figure*}[ht!]
  \begin{center}
      \includegraphics[width=\textwidth]{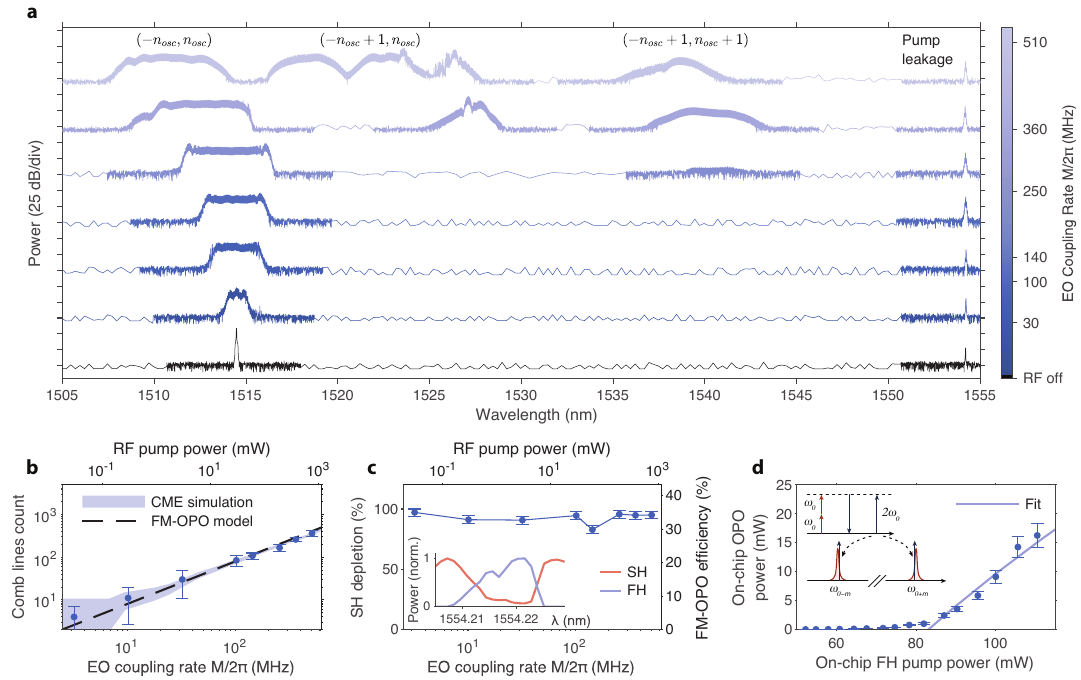}
  \end{center}
 \caption{\textbf{Power-tunability of the FM-OPO. }
\textbf{a},~Combs evolution with the electro-optic drive. As the electro-optic coupling increases, we observe comb growth and the emergence of secondary combs that couple $(-n_{osc}+m, n_{osc})$ modes, aligning with multiple gray lines in Fig.~\ref{fig:fig3}. We combine high-resolution (approximately 20 pm) measurements from areas with detectable signals and low-resolution readings from the rest of the range. We primarily focus on the signal around 1515~nm for simplicity. The peak around 1554~nm is due to a minor leakage of the initial pump into the cavity. 
\textbf{b},~The maximum count of individual oscillating lines in the FM-OPO as a function of electro-optic coupling. The faint blue region corresponds to the prediction from the coupled-mode theory. We plot the comb count predicted with the FM-OPO theory with a dashed line.
\textbf{c},~We measure the FM-OPO optical pump's depletion of around $93\%$, corresponding to the conversion efficiency of around 34$\%$. The inset shows the normalized SHG pump (orange) and FM-OPO (blue) when changing the pump wavelength at $M/2\pi\approx30~\text{MHz}$. The contrast between the SH maximum and minimum defines depletion.
\textbf{d},~The pure OPO's output power as a function of the pump power at the fundamental frequency. The pump initially generates a second harmonic in a single-pass waveguide that later drives the OPO. A line fitting to the data reveals a nonlinear coupling rate $g/2\pi$ of approximately 12~kHz. We describe the displayed measurement and simulation uncertainties in Methods.
}
 \label{fig:fig4}
\end{figure*}

{\bf
Optical frequency combs have revolutionized precision measurement, time-keeping, and molecular spectroscopy~\cite{Diddams2020, Coddington2016, Suh2016, Picque2019, Bao2021}. A substantial effort has developed around ``microcombs'': integrating comb-generating technologies into compact, reliable photonic platforms~\cite{Kippenberg2011}. Current approaches for generating these microcombs involve either the electro-optic~\cite{Zhang2019} (EO) or Kerr mechanisms~\cite{Shen2020}. Despite rapid progress, maintaining high efficiency and wide bandwidth remains challenging. Here, we introduce a new class of microcomb -- an integrated optical frequency comb generator that combines electro-optics and parametric amplification to yield a frequency-modulated optical parametric oscillator (FM-OPO). In stark contrast to EO and Kerr combs, the FM-OPO microcomb does not form pulses but maintains operational simplicity and highly efficient pump power utilization with an output resembling a frequency-modulated laser~\cite{Siegman1986}. We outline the working principles of FM-OPO and demonstrate them by fabricating the complete optical system in thin-film lithium niobate (LNOI). We measure pump to comb internal conversion efficiency exceeding $93\%$ ($34\%$ out-coupled) over a nearly flat-top spectral distribution spanning $\approx 1,000$ modes ($\approx 6~\text{THz}$). Compared to an EO comb, the cavity dispersion rather than loss determines the FM-OPO bandwidth, enabling broadband combs with a smaller RF modulation power. The FM-OPO microcomb, with its robust operational dynamics, high efficiency, and large bandwidth, contributes a new approach to the field of microcombs and promises to herald an era of miniaturized precision measurement, and spectroscopy tools to accelerate advancements in metrology, spectroscopy, telecommunications, sensing, and computing. 
}

\section*{Introduction}

Optical frequency combs, characterized by their precisely spaced, sharp spectral lines that serve as a ``frequency ruler" for light, are indispensable tools in numerous fields, from precision metrology and atomic clocks to high-capacity telecommunications and molecular spectroscopy~\cite{Diddams2020, Coddington2016, Suh2016, Picque2019, Bao2021}. Fueled by their potential practical applications, the drive to miniaturize frequency combs into chip-scale integrated devices, known as microcombs, has recently accelerated at a remarkable pace~\cite{Kippenberg2011}. Traditional optical frequency combs, produced through mode-locked lasers and synchronously pumped optical parametric oscillators, are large-scale and require substantial infrastructure, thus limiting their utility outside laboratory settings. Two principal methods for creating integrated frequency comb sources suitable for smaller, deployable devices have been explored in response. The first involves third-order $\chi^{(3)}$ or Kerr optical nonlinearity, with successful demonstrations in materials such as silica, silicon nitride, aluminum nitride, silicon carbide, and lithium niobate \cite{Okawachi2011, Brasch2016, Jung2013, Guidry2021, Wang2019}. The second strategy employs the electro-optic effect, which has been realized in resonant (shown in Fig.~\ref{fig:fig1}a) and non-resonant integrated thin-film lithium niobate devices \cite{Zhang2019, Hu2022, Yu2022}.
Despite these remarkable advances, electro-optic and Kerr combs face several challenges. They are often limited in their efficiency, exhibit a strong pump background, suffer from limited tunability, and display a decreasing comb line intensity for the lines distant from the pump. Moreover, Kerr frequency combs demand sophisticated control and become significantly more challenging to operate at a smaller free spectral range (FSR).

In this study, we propose and demonstrate a new type of microcomb that combines the advantages of both EO and Kerr combs, merging nonlinear optical processes with electro-optic modulation in an integrated device. Specifically, our structure accommodates both optical parametric amplification and phase modulation within a single cavity, thereby facilitating the generation of a frequency-modulated optical parametric oscillator (FM-OPO, Fig.~\ref{fig:fig1}b).~\cite{Diddams1999, EstebanMartin2012} Remarkably, unlike in conventional Kerr and EO combs, the dynamics in our system \emph{do not result in pulse formation}, making the output more closely resemble that of a frequency-modulated (FM) laser. This strategy maintains the operational simplicity characteristic of electro-optic combs while achieving substantially broader bandwidths than those attainable through modulation alone. Furthermore, our technique gives rise to a flat-top output comb, an optimal spectral distribution for many applications, while avoiding unwanted  nonlinearities that manifest at large pulse peak powers\cite{Okawachi2017}. Finally, the FM-OPO exhibits impressive efficiency, converting a significant fraction of the pump light into comb lines while demanding only modest RF power inputs for operation.

To implement the integrated FM-OPO, we turn to thin-film lithium niobate (LN) for its strong second-order optical nonlinearity and electro-optic (EO) effect. Thin-film LN has recently emerged as a platform for integrated nanophotonics\cite{Zhu2021, Boes2023} through demonstrations of efficient electro-optic modulators \cite{Wang2018, Li2020}, electro-optic combs\cite{Zhang2019, Hu2022}, periodically poled lithium niobate (PPLN) waveguides for frequency conversion\cite{Jankowski2020, Mishra2021, Park2022}, quantum light generation\cite{Kashiwazaki2020, Nehra2022}, resonant second harmonic generation and optical parametric oscillators\cite{Lu2021, McKenna2022, Ledezma2022}, and integration with complex photonic integrated circuits for applications such as laser control\cite{Li2022} and quantum measurements\cite{Stokowski2023}. The above demonstrations are either based on the EO effect that transfers energy between optical modes separated by the RF frequency or the $\chi^{(2)}$ nonlinearity that can provide broadband gain. Combining these two distinct capabilities forms the foundation for the integrated FM-OPO.

\section*{Comb Dynamics}

Both Kerr and EO comb generation fundamentally rely on mode-locking, which subsequently leads to the formation of pulses. However, this process inherently introduces a strong frequency-dependent variation in the intensity of the comb lines that decay exponentially with their offset from the center.
Another considerable challenge posed by pulse formation is the inefficient utilization of pump power, as a continuous wave (CW) pump only overlaps with a small part of the circulating field. Recent advancements have started to address this issue, mainly by exploiting auxiliary resonances \cite{Hu2022, Xue2019} and utilizing pulsed pumps~\cite{Li2022a}. Finally, pulse formation leads to large intracavity peak powers that can engage other unwanted nonlinearities and make comb formation challenging in integrated platforms \cite{Okawachi2017}. We discover here that incorporating parametric gain into an EO-modulated cavity leads to a frequency comb without necessitating pulse formation. Despite the modulation being close to the cavity resonance mode spacing, our system's dynamics strikingly resemble those of an FM laser~\cite{Harris1964, Kuizenga1970, Siegman1986}. As in an FM laser, we will see that the optical frequency of the signal is swept across a bandwidth $\text{B.W.}$ at the rate of the RF modulation $\Omega$.

We first consider the situation without any modulation. We assume that we operate the OPO nondegenerately so that it emits signal and idler tones at mode number offsets $\pm n_\text{osc}$ from a central mode with frequency $\omega_0$ close to $\omega_p/2$. As we introduce RF modulation at frequency $\Omega$ characterized by a mode coupling rate $M$, these signal and idler tones are simultaneously subject to gain and modulation. The pairing of these effects around the signal and idler creates conditions that mirror the dynamics of an FM laser, where phase-insensitive gain and modulation coexist.

In an FM laser, the limiting behavior that prevents mode-locking arises from a detuning between the cavity's FSR and the drive frequency $\Omega$. The FM laser then transitions to chaotic and mode-locked states as this detuning is reduced and the bandwidth is increased to approach the gain bandwidth of the medium or a limit set by the cavity dispersion~\cite{Siegman1986}. The oscillation bandwidth of the FM-OPO is limited by the cavity's dispersion, characterized by mode frequencies \mbox{$\omega_n = \omega_0 + \zeta_1 \times n + \zeta_2/2 \times n^2$}, where $\zeta_1$ and $\zeta_2$ are the cavity FSR near $\omega_0$ and the second-order dispersion, respectively. Under the regime considered, our device avoids the transition to mode-locking behavior. The signal and idler modes are far separated and experience local FSRs near $\pm n_\text{osc}$ that differ from each other by $2 n_\text{osc}\zeta_2$. Moreover, the parametric nature of the process necessitates the simultaneous formation of combs at both signal and idler frequencies. Therefore in the assumed nondegenerate regime, there is always effectively a drive detuning when we consider both signal and idler combs. This results in dynamics that closely mirror those of an FM laser with detuned driving, where continuous frequency sweeping is observed rather than pulse formation. The effective bandwidth is given by
\begin{equation}
\text{B.W.}\equiv2\Gamma\Omega = \frac{4M\Omega}{n_\text{osc}\zeta_2}
\label{eqn:BW}
\end{equation}
where $\Gamma$ is the modulation index, and the signal and idler tones are frequency modulated as $a_\text{s,i}(t)~\approx~A_\text{s,i}e^{-i\omega_\text{i} t}e^{\mp i\Gamma \sin(\Omega t)} e^{i\omega_\text{p}t/2}$. The bandwidth formula aligns well with the established expression for the FM laser bandwidth $\text{B.W.} \propto {M \Omega }/{(\Omega-\text{FSR})}$~\cite{Siegman1986}, with the correspondence being that the FM laser detuning $\Omega-\text{FSR}$ is replaced by the detuning $n_\text{osc}\zeta_2$ between the drive and local FSR  in the FM-OPO. Finally, we note that there are conditions where the above analysis no longer holds, \textit{e.g.}, at (near-)degenerate OPO operation leading to smaller $n_\text{osc}$, at significantly larger $M$, or for dispersion-engineered waveguides that may match the local signal/idler FSRs. Bulk phase-modulated OPOs have already been demonstrated~\cite{Diddams1999, EstebanMartin2012}. We leave the engineering and study of the dynamics of integrated phase-modulated OPOs in a wider set of operating regimes to future work.

\section*{Results}

We demonstrate an optical frequency comb generator based on an FM-OPO integrated on a chip (Fig.~\ref{fig:fig1}c). The device evenly distributes 11~mW of optical power over 200 comb lines using $140~\text{mW}$ of C-band optical pump power and $200~\text{mW}$ of RF modulation power. Comb lines are spaced by about $5.8~\text{GHz}$. We base our device on a racetrack resonator in thin-film lithium niobate on insulator (LNOI) with intrinsic quality factors of around ${Q}_{\mathrm{i}} \approx 10^6$. This resonator holds within it an electro-optic modulator, an optical parametric amplifier, and a high-efficiency wavelength-selective coupler that nearly fully transmits the 780~nm pump while keeping the C-band excitation within the cavity. Figure~\ref{fig:fig2}a shows a schematic design of the device, while Fig.~\ref{fig:fig2}b shows a microscope image of a single FM-OPO device. The coupler allows our device to operate as a doubly resonant OPO where the pump passes through the OPA but is non-resonant in the cavity. One straight section has gold electrodes patterned next to it, enabling electro-optic modulation of the cavity (see the left inset in Fig.~\ref{fig:fig2}b). The other straight section of the cavity is a periodically poled lithium niobate (PPLN) waveguide that provides parametric gain when pumped with the second harmonic (see the right inset in Fig.~\ref{fig:fig2}b for a second harmonic microscope picture of the poled thin-film lithium niobate). In the Methods section, we describe the design and characterization of the waveguides and cavity in detail. 

We generate the 780~nm pump on the same chip in a separate PPLN waveguide. We filter out the original pump field through three on-chip filters of the same design as the intracavity coupler. The high SHG efficiency allows us to achieve considerable optical pump powers using only a standard commercial C-band laser. Figure~\ref{fig:fig2}c shows an example FM-OPO output spectrum when the device is pumped with around $140~\text{mW}$ of FH optical power (corresponding to around $100~\text{mW}$ of SH power) and $200~\text{mW}$ of RF power, equivalent to about 4.5 V peak voltage. We plot an electro-optic comb generated using the same RF power within the same cavity in gray for comparison. We observe a flat comb formation around signal and idler wavelengths and no significant background from the pump. The measured output aligns with our coupled-mode theory model (thick dark blue line) described below. The bottom right inset in Fig.~\ref{fig:fig2}c shows individual lines in a flat spectrum spaced by around 5.8~GHz. The top right inset in Fig.~\ref{fig:fig2}c shows the result of collecting the output using a fast photodetector and an RF spectrum analyzer. In the RF spectrum, we observe narrow lines spaced by the multiples of the cavity FSR, resulting from the FM-OPO sweeping over a frequency-dependent output coupler (see Methods for details).

We can understand nearly all of the salient features of the observed spectra in the context of an approximate time-domain coupled-mode theory analysis. We also use this formulation to derive the formula for the comb bandwidth shown in equation~\ref{eqn:BW}, which agrees well with observations~\ref{fig:fig4}b. We define mode amplitudes $a_n$ to represent the field amplitudes for the $n$-th mode around the fundamental frequency, where $n$ = 0 corresponds to the fundamental mode closest to half of the pump frequency. In this context, $b$ represents the amplitude of the second harmonic pump field. Each mode $n$ has a natural frequency given by the cavity dispersion with $\zeta_1/2\pi \approx 5.8~\text{GHz}$ and $\zeta_2/2\pi \approx 11~\text{kHz}$ corresponding to the cavity FSR and the second-order dispersion, respectively. Other key parameters include the laser drive detuning $\Delta\equiv\omega_{p}/2 - \omega_0$, and the RF drive detuning from the FSR $\delta \equiv \Omega - \zeta_1$. The mode coupling due to modulation $M$, which is proportional to the RF drive voltage, and the nonlinear coupling rate $g$  provide the critical ingredients for realizing the comb dynamics. We also include the loss rates of the considered field amplitudes, $\kappa_{a,n}$ and $\kappa_{b}$. The rate $\kappa_{b}$ corresponds to that of an extremely lossy single-pass ``cavity'' and allows us to approximate our DRO in this coupled-mode theory formulation. We derive all of the model parameters from independent simulations, as well as experimental and theoretical analysis (refer to the Methods section and SI for more details). 
%Finally, the system is driven by the second harmonic with frequency $\omega_{p}$ and power $P_{2\omega}$. 
The resulting  coupled-mode equations are 
\begin{eqnarray}\nonumber
\dot{{a}}_n
=
&\Bigg[&
i\Bigg(
\Delta + n\delta - \cfrac{n^2\zeta_2}{2}
\Bigg)
-\cfrac{\kappa_{a,n}}{2}\, 
\Bigg]
{a}_{n}
\\ 
&-& i M
\Big(
{a}_{n-1} 
+
{a}_{n+1} 
\Big)
- 2ig
{a}_{-n}^{\ast} {b} 
\label{eq:en_motion_main1}
\\
\dot{{b}} 
=
&-& \cfrac{\kappa_b}{2}\,{b} 
-i g \sum_{n} {a}_{n} {a}_{-n}
+
i\sqrt{\kappa_b} \beta_{in}.
\label{eq:en_motion_main2}
\end{eqnarray}
There are two main approximations in these equations. First, we represent the pump field as the excitation of a very lossy mode $b$ -- solutions involving significant spatial variations of the pump field along the waveguide cannot be represented accurately by this model. Secondly, we only include coupling between modes $n$ and $-n$ -- we ignore the weaker coupling between modes with nearby $n$ numbers. For example, coupling between $n$ and $-n+1$ can be present and may become stronger as a function of pump wavelength. Tuning the pump wavelength and consequently the detuning $\Delta$ over a cavity FSR changes the mode pairs that are amplified (see Fig~\ref{fig:fig3}a). Device parameters are summarized in Extended Data Table \ref{table:tab1}).

We tune the output wavelength in the FM-OPO through small adjustments to the pump wavelength, allowing the output to span the full range of the gain spectrum. This tuning is predominantly influenced by the cavity dispersion, mirroring the characteristics observed in an unmodulated OPO \cite{Eckardt1991, McKenna2022}. We show the OPO tuning behavior in Fig.~\ref{fig:fig3}a. The blue traces correspond to measurements with an optical spectrum analyzer (OSA), whereas the gray lines present the predicted tuning behavior based on the waveguide dispersion. The FM-OPO exhibits a similar tuning pattern, as shown in Fig.~\ref{fig:fig3}b. Here, the comb clusters closely follow the expected tuning. By adjusting the pump wavelength by 20~pm, which equates to half of the cavity's free spectral range (FSR), we can access bandwidth of approximately 70~nm for both FM-OPO and OPO.

We measure the spectra generated by the FM-OPO using an optical spectrum analyzer. We find that the device operates continuously and robustly in a nondegenerate mode at around $n_{osc}\approx 800$. In this regime, we expect Eqn.~(\ref{eqn:BW}) to hold to high accuracy. We pump the device at $1554~\text{nm}$ with about $140~\text{mW}$. We step the electro-optic coupling rate of the 5.8-GHz EO modulation between $0$ and around $510~\text{MHz}$ by varying the RF power supplied to the chip. As shown in Fig.~\ref{fig:fig4}a, we observe a frequency comb develop. A number of additional comb clusters labeled $(-n_{osc}+1,n_{osc})$ and $(-n_{osc}+1, n_{osc}+1)$ appear at a drive exceeding $M/2\pi \approx 360~\text{MHz}$; these are described in more detail in the Methods section. We only plot the signal combs (blue detuned) and omit the idler combs (red detuned) for clarity; we provide full spectra in Extended Data Fig.~\ref{fig:figSI8}. The measured spectral peak at around $1554~\text{nm}$ corresponds to a slight leakage of the original FH pump into the cavity. We count the number of generated lines within the 3~dB bandwidth of the flat-top and plot this in Fig.~\ref{fig:fig4}b. We observe good agreement between the data, numerical solution of the coupled-mode equations \ref{eq:en_motion_main1}-\ref{eq:en_motion_main2} (blue shaded region), and the analytical expression for the FM-OPO given by equation \ref{eqn:BW} (dashed line). At the highest EO modulation rate of around $1.2~\text{W}$, we observe over $1,000$ comb lines oscillating together within $-30~\text{dB}$ from the flat-top mean power (see Extended Data Fig.~\ref{fig:figSI8}e for the full spectrum).

The FM-OPO  operates with high efficiency, converting around $34\%$ of the input SH light into comb lines. First, the intracavity conversion efficiency is high, exceeding $90\%$, based on the pump depletion measurement in Fig.~\ref{fig:fig4}c. We calculate it based on the contrast between the measured maxima and minima of the normalized SH power, visible when tuning the pump wavelength, as shown in the inset. Next, the intracavity comb is outcoupled with the cavity escape efficiency $\eta_a \approx 0.36$, which limits the total efficiency of our device. Note that the depletion and the conversion efficiency do not depend on the RF drive strength. The output power of the FM-OPO resembles a typical behavior of an unmodulated OPO in Fig.~\ref{fig:fig4}d, where we observe a threshold of about 47~mW SH power and nonlinear coupling rate $g/2\pi \approx$~12~kHz, lower than the predicted 67~kHz, which we attribute to operating at non-perfect phase matching $\Delta k \neq 0$.

\section*{Discussion}

We have successfully demonstrated a new type of integrated comb generator and established its fundamental operating principles. Our device demonstrates exceptional brightness, flatness, and efficiency while retaining robust operational dynamics. Given that our initial demonstration still has the potential for significant improvements in optical bandwidth by dispersion engineering, RF power consumption by resonant enhancement, and optical conversion efficiency by improved out-coupling, this breakthrough opens the door to a new class of deployable optical frequency combs. For the well-established application of these combs to the problems of spectroscopy, the versatility of the LN material platform allows for spectral coverage from blue light~\cite{Celik2022} into the mid-infrared~\cite{Mishra2021, Mishra2022}, enabling their use in fields such as medical diagnostics \cite{Sordillo2021},  process control in agriculture, food production, and various industrial sectors \cite{Willer2006, Goldenstein2017}. Moreover, the potential of these devices as a source of flat-top combs makes them invaluable for applications from fiber communication systems to FMCW LiDAR~\cite{Martin2018}.

\bibliography{2023_EOOPO}
\clearpage

\section*{Methods}

\subsection*{Device design and Fabrication}

We design our waveguide geometry to maximize the normalized efficiency and interaction rate. Extended Data Figure~\ref{fig:figSI1}a shows a schematic of the periodically poled, X-cut LN waveguide. We chose the ridge height h = 300~nm, slab thickness s = 200~nm, top width w = 1.2 µm, and SiO$_2$ cladding thickness c = 700~nm. We find the guided modes by numerically solving Maxwell's equations with a finite-element solver (COMSOL). Extended Data Figure~\ref{fig:figSI1}a shows the E$_{\mathrm{x}}$ field distribution for a mode at 1550~nm. Extended Data Figure~\ref{fig:figSI1}b presents the bands of the effective index as a function of wavelength in our waveguide geometry. The blue line highlights the fundamental TE mode we use in our nonlinear waveguide and electro-optic modulator. The difference between the effective index at the fundamental and second harmonic frequency $\Delta \mathrm{n}_{\mathrm{eff}}$ results in phase mismatch that we compensate for with periodic poling with a period of around \mbox{$\Lambda = \lambda_{SH}/\Delta \mathrm{n}_{\mathrm{eff}} =$ 3.7 µm}. The LN waveguide forms a racetrack resonator with an intracavity directional coupler designed to close the resonator for the FH but ensure that the SH pump does not circulate. We call this design a ``snail resonator". All of the waveguide bends are defined by Euler curves to minimize light scattering between straight and bent waveguide sections.

We periodically pole the thin-film LN before the waveguide fabrication by patterning chromium finger electrodes on top of an insulating SiO$_2$ layer. Extended Data Figure~\ref{fig:figSI1}c shows an SEM micrograph of a poling electrode. Next, we apply short pulses on the order of 1~kV to invert the ferroelectric domains and then verify the poling with a second harmonic microscope; Extended Data Fig.~\ref{fig:figSI1}d shows a periodically poled film. In the second harmonic microscope picture, the black areas on the sides of the image correspond to the metal electrodes. The oblong shapes stretching between fingers correspond to the inverted LN domains. White regions at the center of the inverted domains correspond to the poling that extends throughout the full depth of the thin-film LN. We pattern the critical \mbox{waveguides} within the fully poled film regions by aligning the electron-beam lithography mask in the waveguide patterning step.

Extended Data Fig.~\ref{fig:figSI2} presents the fabrication process flow. We start with a thin-film lithium niobate on insulator chip (Extended Data Fig.~\ref{fig:figSI2}a). We use 500~nm LN film bonded to around 2 µm of SiO$_2$ on a silicon handle wafer (LNOI from NanoLN). Then, we deposit about 100~nm of silicon dioxide using plasma-enhanced chemical vapor deposition (PlasmaTherm Shuttlelock PECVD System), which serves as a protective layer and prevents leakage current during poling. We pattern 100~nm thick chromium electrodes (evaporated with Kurt J. Lesker e-beam evaporator) on top of the insulating layer through electron-beam lithography (JEOL~6300-FS,~100-kV) and liftoff process and apply short voltage pulses to invert the LN domains (Extended Data Fig.~\ref{fig:figSI2}b). Next, we remove the chromium and SiO$_2$ layers with chromium etchant and buffered oxide etchant to obtain a poled thin-film LN chip (Extended Data Fig.~\ref{fig:figSI2}c). We follow with waveguide patterning using JEOL~6300-FS electron-beam lithography and hydrogen silsesquioxane mask (FOx-16). We transfer the mask to the LN material using dry etching with an argon ion mill (Extended Data Fig.~\ref{fig:figSI2}d). After the waveguide fabrication, we pattern another liftoff mask with electron-beam lithography to pattern electrodes for our electro-optic modulators (Extended Data Fig.~\ref{fig:figSI2}e). We use 200~nm of gold with a 15~nm chromium adhesion layer evaporated with the e-beam evaporator. We clad the entire chip with a layer of 700~nm thick SiO$_2$ deposited with a high-density plasma chemical vapor deposition using PlasmaTherm Versaline HDP CVD System (Extended Data Fig.~\ref{fig:figSI2}f) and open vias to access electrodes using inductively coupled plasma reactive ion etching (Extended Data Fig.~\ref{fig:figSI2}g). We finish preparing the chip facets for light coupling by stealth dicing with a DISCO DFL7340 laser saw.

\subsection*{Experimental Setup}

We characterize our devices' FM-OPO and OPO response using the setup in Extended Data Fig.~\ref{fig:figSI3}. We color-code the paths intended to use with various signals: light orange corresponds to the fundamental harmonic light (around 1500-1600~nm), the blue path corresponds to the second harmonic (around 750-800~nm), and green corresponds to the RF signals. We drive our devices with a tunable C-band laser (Santec TSL-550, 1480–1630~nm) that we amplify with an erbium-doped fiber amplifier (EDFA) to around 1 watt. The wavelength of the laser is controlled in a feedback loop using a wavelength meter (Bristol Instruments 621B-NIR). We control the optical power to the chip with a MEMS variable optical attenuator (from OZ Optics) and calibrate the power using a 5$\%$ tap and a power meter (Newport 918D-IR-OD3R). The light then passes through a fiber polarization controller (FPC) and couples to the chip facet through a lensed fiber. We deliver RF signals to the chip through a ground-signal-groud probe (GGB Industries Picoprobe 40A). We use Keysight E8257D PSG Analog Signal Generator as an RF source and amplify it with a high-power
amplifier (Mini-Circuits ZHL-5W-63-S+). We place a circulator before the chip to avoid any reflections into the source and terminate the reflected port after passing it through a 20 dB attenuator.

The generated light is split between two paths with a 1000-nm short-pass dichroic mirror (Thorlabs DMSP1000). The two paths are connected to the InGaAs and Si avalanche photodiodes (Thorlabs APD410A and Thorlabs APD410) to detect the FH and SH power, respectively. VOAs precede both APDs to avoid saturation and increase the dynamic range of the measurements (HP 8156A and Thorlabs FW102C). Part of the FH path splits into an optical spectrum analyzer (Yokogawa AQ6370C) and a fast photodetector (New Focus 1554-B-50), which response is characterized by an RF spectrum analyzer (Rohde $\&$ Schwarz FSW26).

\subsection*{Intracavity coupler characterization}

We characterize the performance of the intracavity coupler using a smaller resonator with a straight section length of around 2 mm. Extended Data Figure~\ref{fig:figSI4}a shows transmission of such a cavity (depicted in Extended Data Fig.~\ref{fig:figSI4}b), where we normalize the background to one. We observe the contrast of cavity modes changing across the used wavelength range due to the changes in the intrinsic and extrinsic quality factors. The former can be used to benchmark the coupler's performance. We observe a smooth transition from an undercoupled cavity at 1500~nm, through critical coupling at around 1550~nm, to an overcoupled cavity at 1580~nm. To verify this, we fit the quality factors of all the modes. An example is shown in Extended Data Fig.~\ref{fig:figSI4}c, where we observe intrinsic quality factor $Q_i \approx \, 2.5 \cdot 10^6$ and extrinsic quality factor $Q_e \approx \, 0.8 \cdot 10^6$. Extended Data Figure~\ref{fig:figSI4}d shows the intrinsic and extrinsic quality factors measured as a function of wavelength. We find that $Q_i$ peaks at around 1580~nm, corresponding to the maximum transmission through the coupler. In the FM-OPO device, we use the same coupler but extend the device length to 10 mm, which results in the flattening of the $Q_i$ dependence on wavelength.

\subsection*{Dispersion measurement} 

The second-order dispersion $\zeta_2$ is a critical parameter of the FM-OPO because it determines the comb span and tunability. To quantify it, we modify the measurement setup by adding another $5\%$ tap connected to a fiber Mach-Zehnder interferometer (MZI) and a photodetector (Newport 1623 Nanosecond Photodetector), see Extended Data Fig.~\ref{fig:figSI5}a. We collect the MZI transmission and the cavity transmission while scanning the pump laser and calibrate the wavelength by unwrapping the phase in the MZI transmission spectrum. This method allows us to measure cavity mode location with precision on the order of single MHz. We measure the FM-OPO cavity spectrum using the feedline waveguide and extract the local FSR, as shown in Extended Data Fig.~\ref{fig:figSI5}b. The relative position of cavity modes is defined by $\omega_n = \omega_0 + \zeta_1 \times n + \zeta_2/2 \times n^2$. We fit the FSR with respect to the mode number and extract the second-order dispersion parameter $\zeta_2/2\pi~\approx$~11~kHz, which agrees with the theoretical prediction based on the finite-element simulation.
 
\subsection*{Second-order optical nonlinearity characterization}

We characterize the nonlinear performance of our PPLN waveguides through a second harmonic generation measurement in a waveguide that passes through the same poled area of the chip as the FM-OPO PPLN waveguides. The experiment geometry is shown in Extended Data Fig.~\ref{fig:figSI6}a, where the input to the chip is the same as in the general setup but the lensed fiber couples to the test waveguide. Two APDs collect the output light the same way as in the FM-OPO measurements. Extended Data Figure~\ref{fig:figSI6}b shows an example of a measured SHG transfer function recorded while sweeping the C-band laser with fixed power of around 200 µW on the chip. The waveguide length is about 7 mm, and slight distortion to the sinc function results from small waveguide nonuniformities along its length. Extended Data Figure~\ref{fig:figSI6}c shows the peak SH power on-chip recorded as a function of the on-chip pump power at the FH frequency. The inset shows a bright SH spot scattered at the end of an on-chip LN waveguide and lensed fiber tip. We fit a quadratic polynomial to the data to extract the normalized efficiency $\eta$ that defines the relationship between the SH and pump power:
\begin{eqnarray}
{P}_{\mathrm{SH}} = \eta {P}_{\mathrm{FH}}^2 \, L^2,
\label{eq:SHG}
\end{eqnarray}
where $L$ is the length of the PPLN waveguide, ${P}_{\mathrm{SH}}$ and ${P}_{\mathrm{FH}}$ correspond to the power of the second harmonic and fundamental, respectively. We extract normalized efficiency of around 1,500 $\%/(\mathrm{Wcm}^2)$, corresponding to the interaction rate around $g/2\pi \approx $ 67 kHz, which agrees with our theory. The measured FM-OPO operates away from the perfect quasi-phase matching, $\Delta k \neq 0$, which reduces the interaction rate to around 12~kHz.

\subsection*{Electro-optic characterization}

To characterize the electro-optic performance of the FM-OPO resonator, we drive the cavity with RF and probe the transmission spectra of the feedline waveguide as shown in Extended Data Fig.~\ref{fig:figSI7}a. We use the same input chain as in the FM-OPO measurements, except for the RF amplifier. We collect the light using an InGaAs APD paired with a VOA. The cavity transmission with no RF drive reveals a usual Lorentzian lineshape (blue points in Extended Data Fig.~\ref{fig:figSI7}b), that we fit to extract the intrinsic quality factor of around ${Q}_{i} \approx\, 1\cdot 10^6$. However, the lineshape becomes distorted when the RF modulation is applied to the cavity on resonance with the local FSR. We model it by simplifying the full FM-OPO cavity coupled-mode equation \ref{eq:en_motion_main1} and adding an FH drive to one of the cavity modes $n = 0$. In the small optical power limit and absence of the SH drive, we can write the model as:
\begin{eqnarray}\nonumber
\dot{{a}}_n
&=&
\left( i\left(\Delta - \cfrac{ n^2 \zeta_2}{2} \right)
- \cfrac{\kappa_{a,n}}{2}
\right) {a}_{n}
\\
&-& i M
\left(
{a}_{n-1}
+
{a}_{n+1} \right)
+ i \sqrt{\cfrac{\kappa_a^{(e)} P_{\mathrm{FH}}}{\hbar \omega_a}}\, \delta_{n,0}.
\label{eq:eo_eqns_of_motion}
\end{eqnarray}
Here, $\Delta$ is the laser detuning, and $\delta_{n,0}$ is a Kronecker delta. We model the EO-modulated cavity response by solving this system of equations for 50 modes in steady state:
\begin{eqnarray}
0 = \overline{\overline{\bf{M}}} \, \overline{\bf{A}}\, + \, \overline{\bf{B}},
\label{eq:M_formulation}
\end{eqnarray}
where $\overline{\overline{\bf{M}}}$ is the matrix including pump detuning, loss rates of the cavity modes, and electro-optic coupling, \mbox{$\overline{\bf{B}}(n=0)~=~i~\sqrt{({\kappa_a^{(e)} P_{\mathrm{FH}}})/({\hbar \omega_a}})$} and $\overline{\bf{B}}(n~\neq~0)~=~0$. We find $ \overline{\bf{A}}\, = - \overline{\overline{\bf{M}}} \,^{-1} \, \overline{\bf{B}}$. The total output power of the cavity consists of the laser pump interfering with the intracavity field and a sum of all the generated sidebands:
\begin{eqnarray}
|a_{\mathrm{out}}|^2 = 
\left|\left(a_{\mathrm{in}} - i \sqrt{\kappa_{0}^{\mathrm{(e)}}} a_{0}
\right)\right|^2
+ \sum_{n\neq 0} \kappa_{n}^{\mathrm{(e)}} |a_{n}|^2.
\label{eq:a_out_EO}
\end{eqnarray}
We evaluate this model numerically to fit the transmission lineshapes of the modulated cavity for various peak voltage values. The orange points in Extended Data Fig.~\ref{fig:figSI7}b correspond to one example of data collected for the cavity modulated with a peak voltage of around $V_{\mathrm{P}} \approx $ 4.5 V. The red line corresponds to the fit. When fitting the modulated lineshapes, we fix the extrinsic and intrinsic quality factors, as measured for the unmodulated line, and extract only the electro-optic coupling $M$. Then, we plot the measured values of the EO coupling $M/2\pi$ as a function of peak voltage in Extended Data Fig.~\ref{fig:figSI7}c and fit a line to find the dependence of the EO coupling on the peak voltage. We measure $M/2\pi \approx 60$ MHz/V.

\subsection*{RF and optical spectra of the FM-OPO}

We examine the FM-OPO combs we produce using a high-speed photodetector and an RF spectrum analyzer. Interestingly, a single FM-OPO, as defined by equations (derived in SI):
\begin{eqnarray}
a_\text{i}(t) &=& A_\text{i}e^{-i\omega_\text{i} t}e^{i\Gamma \sin(\Omega t)} e^{i\omega_\text{p}t/2}
\label{eq:FM_sol1}
\\
a_\text{s}(t) &=& A_\text{s}e^{-i\omega_\text{s} t}e^{-i\Gamma \sin(\Omega t)}e^{i\omega_\text{p}t/2},
\label{eq:FM_sol2}
\end{eqnarray}
should not create any detectable RF tones when evaluated with a fast photodetector since a pure phase or frequency modulation will not be detected on a photodiode measuring intensity. However, we observe peaks in the RF spectra for the FM-OPOs shown in Fig.~\ref{fig:fig4}a that are spaced by $\Omega$. These are displayed in Extended Data Fig.~\ref{fig:figSI8}a, and we provide a closer look at the first sidebands in Extended Data Fig.~\ref{fig:figSI8}b.

We find that even a minor dependence of the cavity's external coupler transmission on wavelength can lead to a noticeable conversion from frequency modulation to intensity modulation. To confirm this, we estimate the expected result of a high-speed photodiode measurement of signal and idler combs produced following equations \ref{eq:FM_sol1}-\ref{eq:FM_sol2}, under the influence of a wavelength-dependent coupler. We determine the external coupling as a function of frequency for our cavity from the same measurement we used for dispersion characterization. The average change in the external coupling across the 1500-1600~nm measurement bandwidth is approximately $\partial\kappa_a^{(e)}/\partial \omega \approx -5\cdot10^{-6}$. The calculated RF spectra (Extended Data Fig.~\ref{fig:figSI8}c) qualitatively match our experimental observations, with discrepancies occurring at higher electro-optic modulation rates where the single FM-OPO approximation is no longer applicable. For each RF spectrum, we also present the full optical spectra (including signal and idler) in Extended Data Fig.~\ref{fig:figSI8}d. 

We plot the spectrum with the largest observed coverage, measured at around $1.2~\text{W}$ of RF power in Extended Data Fig.~\ref{fig:figSI8}e. Note that for a particular pump wavelength, there are multiple possible modes of oscillation corresponding to the coupling between different mode pairs $(-n_{osc}, n_{osc})$, $(-n_{osc}-1, n_{osc})$, $(-n_{osc}-1, n_{osc}-1)$, and so on. For the non-modulated OPO operation at the power levels we experimentally characterized, we observe only one oscillating mode at a fixed pump wavelength (i.e., $(-n_{osc}, n_{osc})$), which we attribute to optimal phase matching. Adjusting the pump results in switching between different mode pairs with a periodicity of 1/2 FSR. However, in the presence of sufficiently strong modulation, clusters of modes arise in the FM-OPO spectrum corresponding to these secondary mode pairs being excited.

\subsection*{FM-OPO tuning with laser and RF detuning}

We experimentally analyze the behavior of FM-OPO comb properties with respect to the RF drive parameters. First, we step the pump laser across one FSR of the cavity (Extended Data Fig.~\ref{fig:SI9}a) and record the OSA spectra for various electro-optic coupling rates (Extended Data Fig.~\ref{fig:SI9}b-d). We can calibrate the pump wavelength, as shown in Extended Data Fig.~\ref{fig:SI9}a with respect to the cavity modes by looking at the slight leakage of the original FH pump visible as a faint line at around 1554~nm signal wavelength in all the colormaps. In this study, we operate in a nondegenerate regime and observe a pure OPO in Extended Data Fig.~\ref{fig:SI9}b. Next, by switching on a moderate RF modulation, we achieve $M/2\pi~\approx$~100~MHz in Extended Data Fig.~\ref{fig:SI9}c and observe comb formation and higher-order FM-OPO comb development. Finally, at high modulation of around $M/2\pi~\approx$~510~MHz, we observe that the combs originating from different OPO modes $(-n_{osc}, n_{osc})$, $(-n_{osc}+1, n_{osc})$, and $(-n_{osc}+1, n_{osc}+1)$ start to merge. We note that the areas with suppressed FM-OPO intensity result from the waveguide mode crossings between the fundamental TE mode and higher order modes that effectively reduce the quality factors in that region. Next, we analyze the FM-OPO response to the RF detuning $\delta$, defined schematically in Extended Data Fig.~\ref{fig:SI9}e. We measure this by pumping the device at around 1545~nm and using $M/2\pi \approx$ 510 MHz. For most measurements, we fix the detuning to $\delta = 0$ so that the RF frequency is on resonance with the cavity FSR near degeneracy $\Omega = \zeta_1$ to maximize the comb span and output optical power. If the RF drive is detuned, we observe comb shrinking, as shown in Extended Data Fig.~\ref{fig:SI9}f, and the total output power decreases, as shown in Extended Data Fig.~\ref{fig:SI9}g.

\subsection*{Uncertainty analysis}

The measurement error of the comb count in Fig.~\ref{fig:fig4}b is given by the standard deviation of 51 measurements (41 measurements for the highest RF power). The shaded region corresponds to the coupled-mode-equation simulation, from which we extract the half-widths of the simulated combs. We assume uncertainty of $\pm1$ mode on each side of the signal and idler combs. We calculate the uncertainty of the measured depletion of the SH pump (Fig.~\ref{fig:fig4}c) and the measured OPO signal (Fig.~\ref{fig:fig4}d) based on the standard deviation of the SH and FH signals over the measurement time.

We measure the FM-OPO resonator's average intrinsic and total quality factors by averaging the results of Lorentzian fits over around 20~nm of the spectrum, where we observe the comb formation. The standard deviation gives their uncertainties. We infer the uncertainty of $\kappa_b$ based on the precision of our estimation of the group index ($10^{-3}$, based on the finite-element solver). We calculate the cavity escape efficiency uncertainty based on the errors of the average quality factors. The uncertainties of the cavity free spectral range, cavity dispersion, peak waveguide nonlinear efficiency, and electro-optically induced mode-coupling rate correspond to the standard errors of the fit parameters extracted from the least-square fitting. Uncertainties of the nonlinear interaction rate and the SH power threshold of the OPO are calculated based on the standard errors of a nonlinear fit. We calculate the internal and total OPO efficiency errors based on the cavity escape efficiency and SH depletion uncertainties.

\clearpage
% \bibliography{2023_EOOPO}

\section*{Data availability} 
The data sets generated during and/or analyzed during this study are available from the corresponding author on request.

\section*{Acknowledgements} 
This work was supported by U.S. government through the Defense Advanced Research Projects Agency Young Faculty Award and Director's Fellowship (YFA, Grant No. D19AP00040), LUMOS program (Grant No. HR0011-20-2-0046),  the U.S. Department of Energy (Grant No. DE-AC02-76SF00515) and Q-NEXT NQI Center, and the U.S. Air Force Office of Scientific Research provided a MURI grant (Grant No. FA9550-17-1-0002). We thank NTT Research for their financial and technical support. H.S.S. acknowledges support from the Urbanek Family Fellowship, and V.A. was partially supported by the Stanford Q-Farm Bloch Fellowship Program and the Max Planck Society Sabbatical Fellowship Award. This work was also performed at the Stanford Nano Shared Facilities (SNSF), supported by the National Science Foundation under award ECCS-2026822. We also acknowledge the Q-NEXT DOE NQI Center and the David and Lucille Packard Fellowship for their support. D.D. and A.Y.H acknowledge support from the NSF GRFP (No. DGE-1656518). H.S.S. and V.A. thank Kevin Multani and Christopher Sarabalis for discussions and technical support. A.H.S.-N. thanks Joseph M. Kahn and Stephen E. Harris for useful discussions.

\section*{Author contributions}
A.H.S.-N. and H.S.S. conceived the device and H.S.S. designed the photonic integrated circuit. 
H.S.S., C.L., and M.J. developed essential components of the photonic circuit.
H.S.S., T.P., and A.Y.H. fabricated the device.
H.S.S., V.A., and O.T.C. developed the fabrication process.
M.M.F. and A.H.S.-N. provided experimental and theoretical support.
H.S.S., T.P., and D.J.D. performed the experiments.
H.S.S., A.Y.H., T.P., and D.J.D.analyzed the data.
H.S.S., and A.H.S.-N. wrote the manuscript.  
H.S.S., V.A., and A.H.S.-N. developed the experiment. H.S.S., D.J.D., A.H.S.-N. developed the numerical and analytical models. A.H.S.-N. supervised all efforts. 

\section*{Competing interests}
A.H.S.-N., H.S.S., and A.Y.H. are inventors of a patent application that covers the concept and implementation of the frequency-modulated optical parametric oscillator and its applications. The remaining authors declare no competing interests.

% Extended data figures
\renewcommand{\tablename}{Extended Data Table}
\renewcommand{\figurename}{Extended Data Fig.}
\setcounter{figure}{0}

\begingroup
\setlength{\tabcolsep}{3.5pt} %
\renewcommand{\arraystretch}{1.1} %
\begin{table*}[t]
\begin{tabular}{c|c|l|r|r}
Parameter                  & Units             & \multicolumn{1}{c|}{Description}                                                                                                   & \multicolumn{1}{c|}{Value} & \multicolumn{1}{c}{Uncertainty} \\ \hline
Average $Q_a~(10^6)$       & -                 & \begin{tabular}[c]{@{}l@{}}Average FH quality factor, \\ measured by laser spectroscopy of cavity at FH\end{tabular}              & 1.0                        & 0.2                             \\ [4mm]
Average $Q_a^{(i)}~(10^6)$ & -                 & \begin{tabular}[c]{@{}l@{}}Average FH intrinsic quality factor, \\ measured by laser spectroscopy of cavity at FH\end{tabular}              & 1.6                        & 0.3                             \\ [4mm]
$\kappa_b/(2\pi)$          & GHz               & \begin{tabular}[c]{@{}l@{}}Effective loss rate of the SH pump, \\ calculated from simulated group velocity\end{tabular}            & 3.566                      & 0.008                           \\ [4mm]
$\eta_{a}$                 & -                 & \begin{tabular}[c]{@{}l@{}}Cavity escape efficiency, \\ extracted from average $Q_a^{(e)}$ and $Q_a$\end{tabular}                          & 0.36                       & 0.04                            \\ [4mm]
$\zeta_1/(2\pi)$           & GHz               & \begin{tabular}[c]{@{}l@{}}Cavity free spectral range, \\ measured by laser spectroscopy of cavity at FH\end{tabular}              & 5.7853                     & 0.0002                          \\ [4mm]
$\zeta_2/(2\pi)$           & kHz               & \begin{tabular}[c]{@{}l@{}}Cavity dispersion, \\ measured by laser spectroscopy of cavity at FH\end{tabular}                       & 11.1                       & 0.5                             \\ [4mm]
$\eta$                     & $\%/\text{Wcm}^2$ & \begin{tabular}[c]{@{}l@{}}Peak waveguide nonlinear efficiency, \\ measured on a PPLN waveguide\end{tabular}                       & 1,500                      & 100                             \\ [4mm]
$g$                        & kHz               & \begin{tabular}[c]{@{}l@{}}Nonlinear interaction rate, \\ extracted from fitting output power of the OPO\end{tabular}              & 12                         & 2                               \\ [4mm]
$M_0$                      & MHz/V             & \begin{tabular}[c]{@{}l@{}}Electro-optically induced mode-coupling rate, \\ fitted from modulated cavity transmission\end{tabular} & 60                         & 1                               \\ [4mm]
$P_{\text{th}}$            & mW                & \begin{tabular}[c]{@{}l@{}}SH power threshold of the OPO, \\ extracted from fitting output power of the OPO\end{tabular}           & 47                         & 1                               \\ [4mm]
$\rho_{\text{int}}$        & $\%$              & \begin{tabular}[c]{@{}l@{}}Internal OPO efficiency, \\ inferred from the SH depletion\end{tabular}                                 & 93                         & 3                               \\ [4mm]
$\rho_{\text{max}}$        & $\%$              & \begin{tabular}[c]{@{}l@{}}Total OPO efficiency, \\ inferred from the SH depletion and $\eta_{a}$\end{tabular}                     & 34                         & 4                              
\end{tabular}
\caption{
\textbf{Summary of the FM-OPO device parameters.}
}
\label{table:tab1}
\end{table*}
\endgroup

\begin{figure*}[t]
  \begin{center}
      \includegraphics[width=\textwidth]{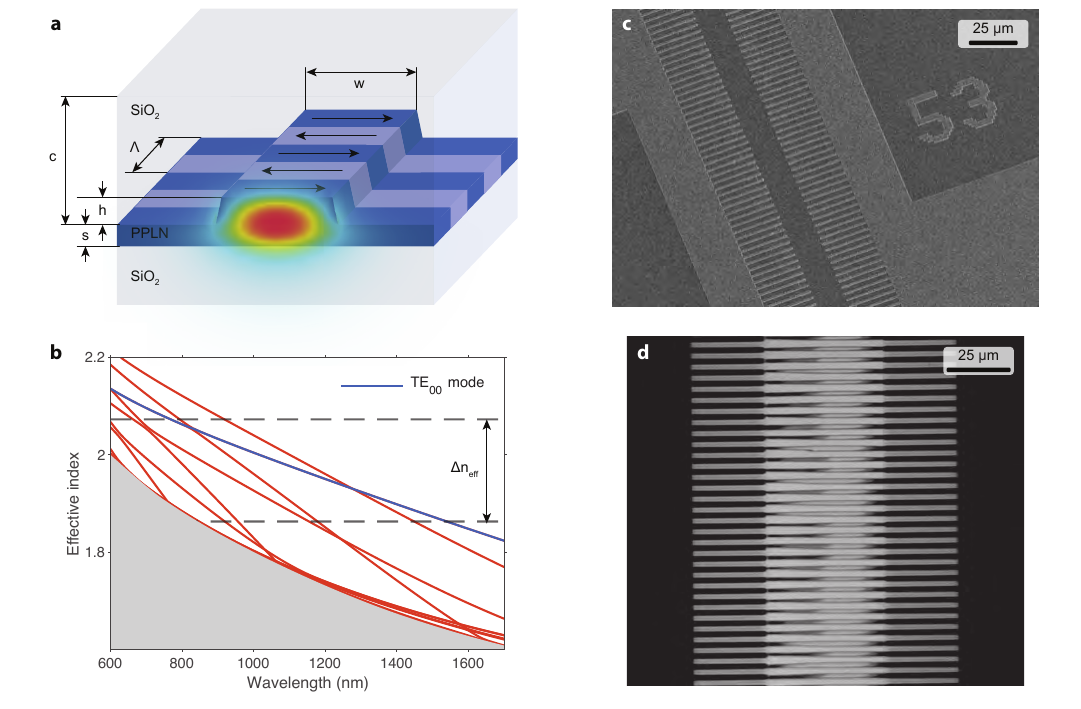}
  \end{center}
 \caption{\textbf{Periodically poled lithium niobate waveguide design. }
\textbf{a},~Schematic of a periodically poled lithium niobate waveguide, waveguide dimensions are: ridge height h = 300~nm, slab height s = 200~nm, width w = 1200~nm, cladding thickness c = 700~nm, poling period $\Lambda$ = 3.7 µm. Thin-film lithium niobate is bonded to a 2 µm thick silicon dioxide layer and clad with a PECVD layer of SiO$_2$. An eigenmode solution at 1550~nm is overlaid with the waveguide schematic.
\textbf{b},~Effective index bands for various waveguide modes in our waveguide geometry. The blue line highlights the fundamental TE mode. We compensate for the effective index mismatch $\Delta \mathrm{n}_{\mathrm{eff}}$ between the fundamental and second harmonic with periodic poling of the film.
\textbf{c},~SEM micrograph of a chromium electrode patterned on a thin-film-LN chip for poling.
\textbf{d},~Second harmonic microscope picture of periodically poled thin-film LN. Black areas on the left and right correspond to Cr electrodes. Oblong, grayscale shapes between the finger electrodes are inverted crystal domains. White areas of the inverted domains correspond to the full-depth poling of the film.
}
 \label{fig:figSI1}
\end{figure*}

\begin{figure*}[t]
  \begin{center}
      \includegraphics[width=\textwidth]{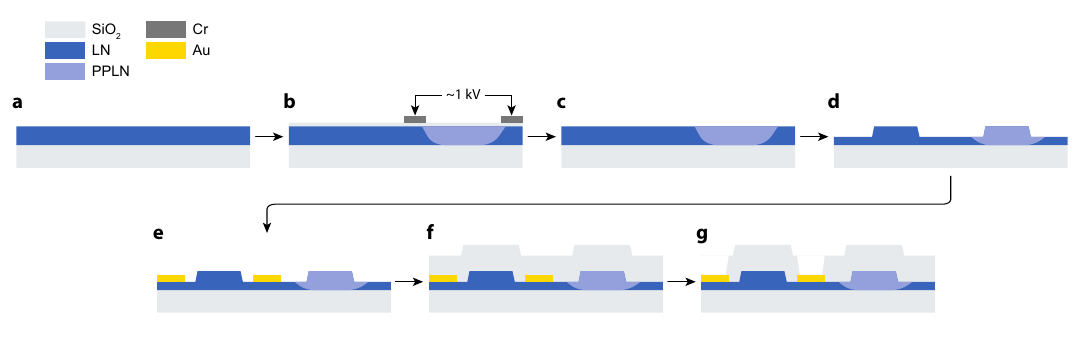}
  \end{center}
 \caption{\textbf{Fabrication process of the photonic integrated circuit. }
\textbf{a},~We start our fabrication process with a thin-film of lithium niobate on insulator (LNOI).
\textbf{b},~Next, we deposit a 100~nm protective layer of SiO$_2$, pattern Cr electrodes, and pole the LN by applying high voltage pulses. 
\textbf{c}-\textbf{d},~We remove the SiO$_2$ and Cr afterward and etch waveguides into the LNOI film through argon ion-mill dry etching.
\textbf{e}-\textbf{f},~After the waveguide fabrication, we pattern gold electrodes with the liftoff process and clad the entire structure with SiO$_2$. 
\textbf{g},~Finally, we pattern vias in the SiO$_2$ layer to access the metal electrodes.
}
 \label{fig:figSI2}
\end{figure*}

\begin{figure*}[t]
  \begin{center}
      \includegraphics[width=\textwidth]{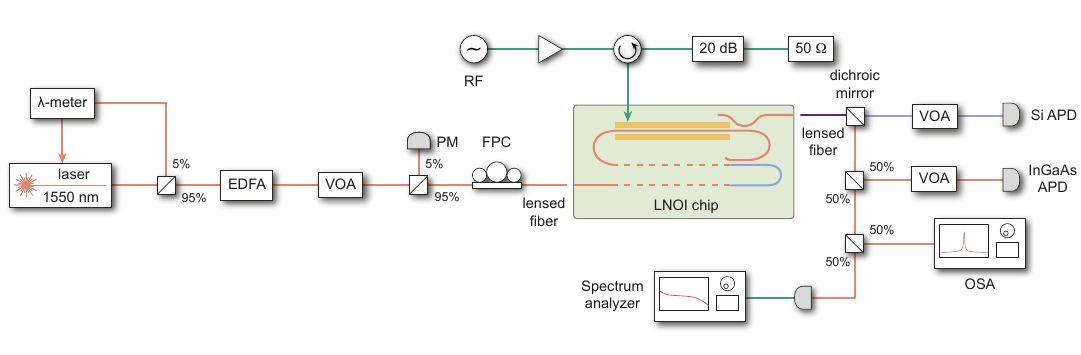}
  \end{center}
 \caption{\textbf{Experimental setup for the FM-OPO characterization. }
We characterize our devices with a C-band tunable laser that we amplify with an erbium-doped fiber amplifier (EDFA), yielding up to 1 W of optical power. We control the power going to the chip with a variable optical attenuator (VOA) and calibrate that power by splitting around 5$\%$ of laser into a power meter (PM). We control the polarization with a fiber polarization controller (FPC) and inject the light into a cleaved chip facet with a lensed fiber. We drive the FM-OPO with an RF source connected to an amplifier. We place a circulator before the chip to avoid reflections returning to the source. Any RF reflections are passed to termination through a 20 dB attenuator. We characterize the output of our devices by splitting the FH and SH light using a dichroic mirror. Both wavelengths are passed through VOSa for power control and measured with calibrated avalanche photodetectors (APDs). Finally, we split part of the FH light into an optical spectrum analyzer (OSA) and a fast photodiode.
}
 \label{fig:figSI3}
\end{figure*}

\begin{figure*}[t]
  \begin{center}
      \includegraphics[width=\textwidth]{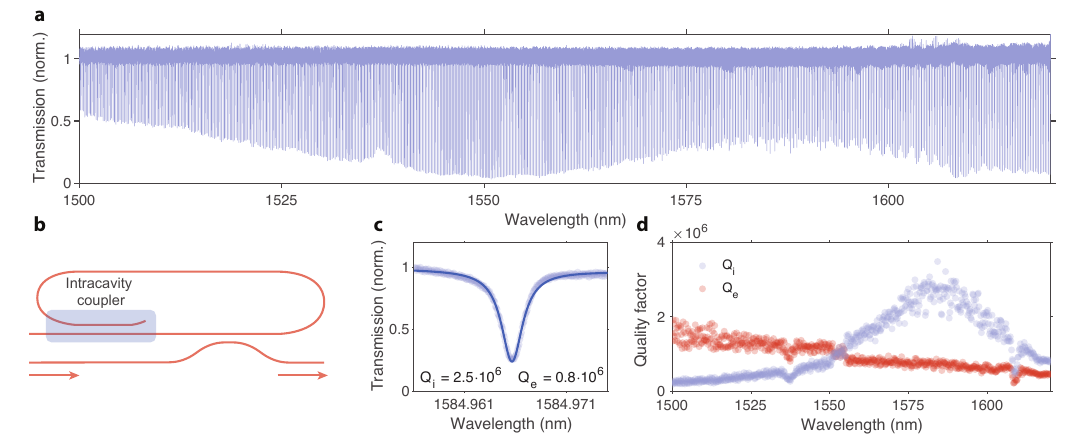}
  \end{center}
 \caption{\textbf{Intracavity coupler characterization. }
\textbf{a},~Broadband transmission spectrum of a snail resonator with a straight section length of 2~mm. We observe mode contrast changing from under-coupled at 1500~nm to critically coupled at around 1550~nm to over-coupled at 1580~nm.
\textbf{b},~Device measurement scheme, we probe a cavity with an internal coupler using an evanescent coupler feed waveguide.
\textbf{c},~Zoom into a single cavity mode around 1585~nm, fitting a Lorentzian lineshape (solid blue line) reveals an intrinsic quality factor ($Q_i$) of around 2.5 million and extrinsic quality factor ($Q_e$) of around 800,000.
\textbf{d},~Intrinsic and extrinsic quality factors as a function of wavelength. We distinguish between the $Q_i$ and $Q_e$ by observing the wavelength dependence. $Q_i$ peaks at around 1580~nm, where the intracavity coupler transmits all the light, thus forming a low-loss cavity. $Q_e$ decreases with wavelength because modes become less confined and can couple stronger to neighboring waveguides. The region where $Q_i \approx Q_e$ around 1552~nm is ambiguous; we report the same average number for both $Q_i$ and $Q_e$ there.
}
 \label{fig:figSI4}
\end{figure*}

\begin{figure*}[t]
  \begin{center}
      \includegraphics[width=\textwidth]{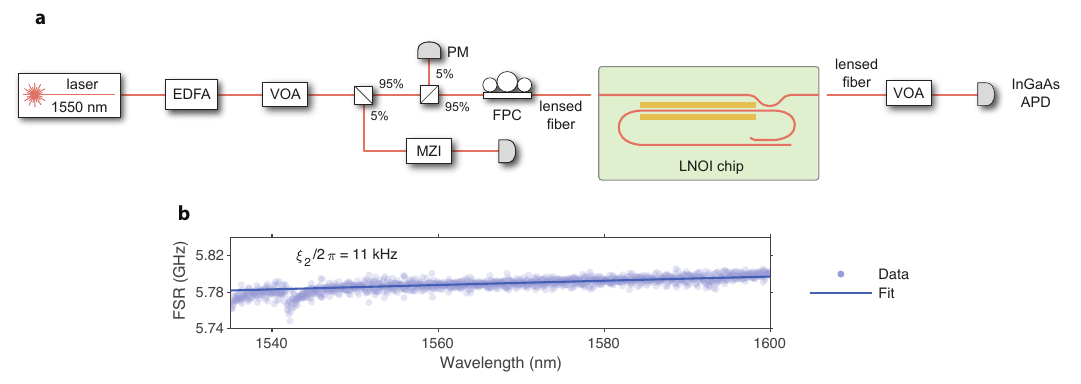}
  \end{center}
 \caption{\textbf{Second-order dispersion characterization of the optical cavity. }
\textbf{a},~We characterize the second-order dispersion by probing the snail cavity with a broadband tunable C-band laser. We use a similar input setup to the one in Fig.~\ref{fig:figSI3} with an additional power splitter connected to a fiber MZI and detector, which serve as wavelength calibration. We collect the light with a setup with a VOA and an InGaAs APD.
\textbf{b},~Measured free spectral range of the cavity as a function of wavelength. We find mode locations and fit a line to extract the second-order dispersion parameter of around $\zeta_2/2\pi \approx$ 11~kHz.
}
 \label{fig:figSI5}
\end{figure*}

\begin{figure*}[t]
  \begin{center}
      \includegraphics[width=\textwidth]{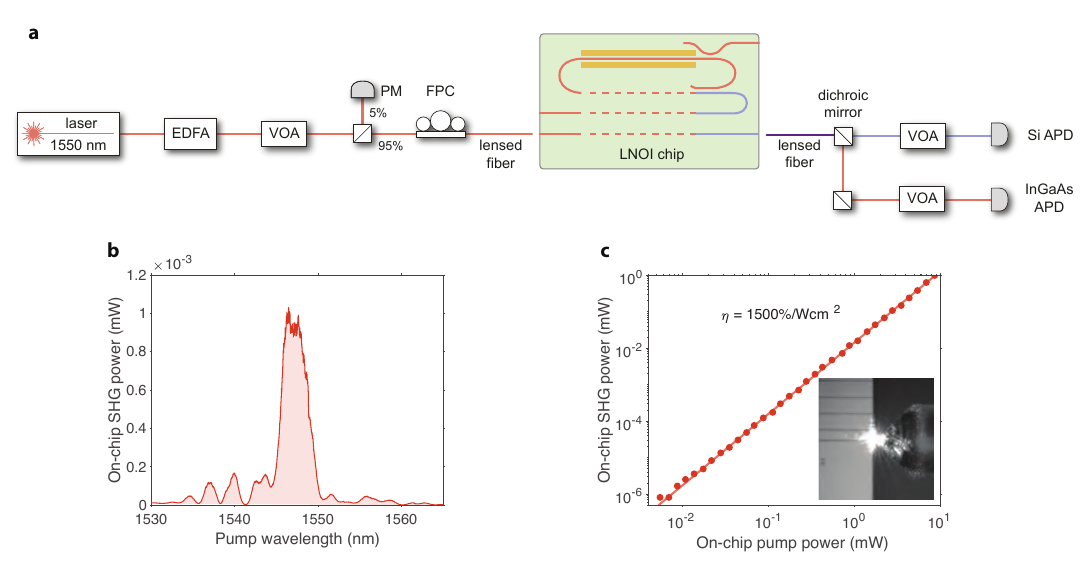}
  \end{center}
 \caption{\textbf{Characterization of the second-order optical nonlinearity. }
\textbf{a},~Experimental setup for the second harmonic generation measurement. We use a similar input setup as in Fig.~\ref{fig:figSI3} and drive a PPLN waveguide with a tunable C-band laser. We collect the output light into a fiber and split it with a dichroic mirror between two avalanche photodetectors (APDs) for FH and SH light characterization.
\textbf{b},~Example SHG transfer function measured at pump power of about $\mathrm{P}_{\mathrm{FH}} \approx $ 200 µW.
\textbf{c},~Measured SHG output power as a function of pump power. The quadratic fit yields a normalized efficiency of around $\eta \approx $ 1,500 $\%/\mathrm{Wcm}^2$. The inset shows a microscope picture of bright SHG light scattered at the output facet of the chip and collected into a lensed fiber.
}
 \label{fig:figSI6}
\end{figure*}

\begin{figure*}[t]
  \begin{center}
      \includegraphics[width=\textwidth]{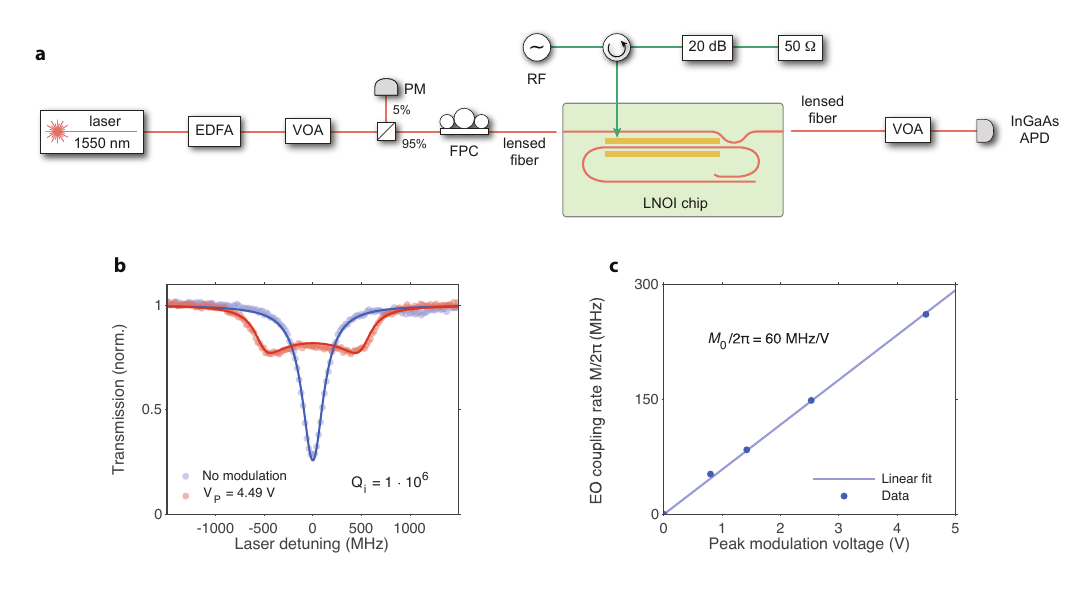}
  \end{center}
 \caption{\textbf{Characterization of the electro-optic coupling to the snail resonator. }
\textbf{a},~Experimental setup. We use a similar input path as in Fig.~\ref{fig:figSI3} and drive the device with a tunable C-band laser. In addition, we deliver RF signals from the source in the same way as in Fig.~\ref{fig:figSI3}, except we do not use the microwave amplifier. We collect the light into an InGaAs avalanche photodiode (APD) through a variable optical attenuator (VOA).
\textbf{b},~Normalized transmission of a single cavity mode with (orange) and without RF drive (blue). Solid lines correspond to fitting the model. We drive the cavity with peak voltage of around $\mathrm{V}_{\mathrm{P}} \approx$ 4.5 V for the modulated dataset.
\textbf{c},~Fitted electro-optic coupling ${M}/2\pi$ as a function of peak modulation voltage. We extract ${M}_0$ from curves like the ones in Fig.~\ref{fig:figSI7}b and fit a line to find ${M}_0/2\pi \approx $ 60 MHz/V.
}
 \label{fig:figSI7}
\end{figure*}

\begin{figure*}[t]
  \begin{center}
      \includegraphics[width=\textwidth]{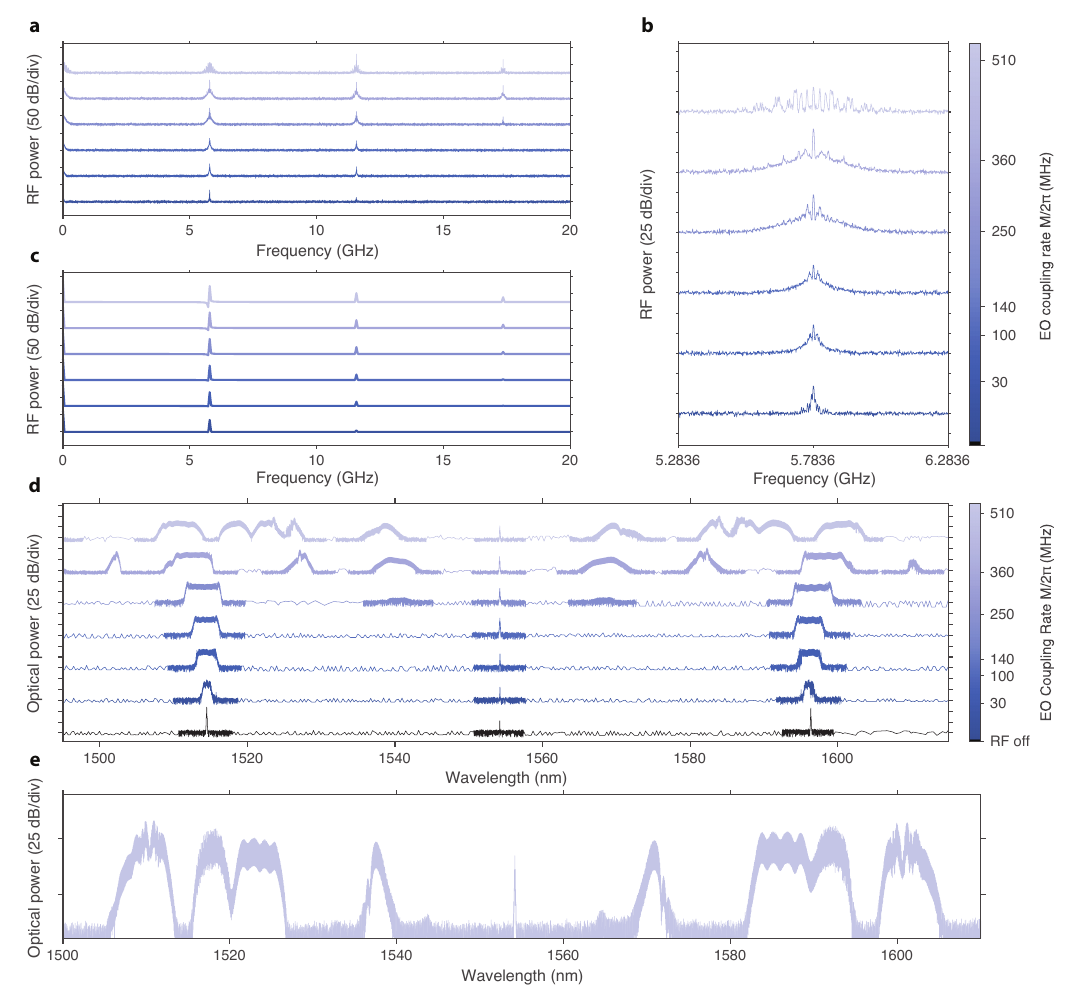}
  \end{center}
 \caption{\textbf{RF and optical spectra generated by the FM-OPO}
\textbf{a},~Measured RF spectra generated by the FM-OPO as a function of the RF drive strength.
\textbf{b},~Zoom-into the first sideband around 5.78~GHz.
\textbf{c},~Simulated RF spectra generated by an FM-OPO coupled to a wavelength-dependent output coupler.
\textbf{d},~Full OSA spectra of the generated FM-OPO combs, partially plotted in Fig.~\ref{fig:fig4}a.
\textbf{e},~Maximum spectral coverage of the FM-OPO, observed with around 1.2~W RF drive power.
}
 \label{fig:figSI8}
\end{figure*}

\begin{figure*}[ht!]
  \begin{center}
      \includegraphics[width=\textwidth]{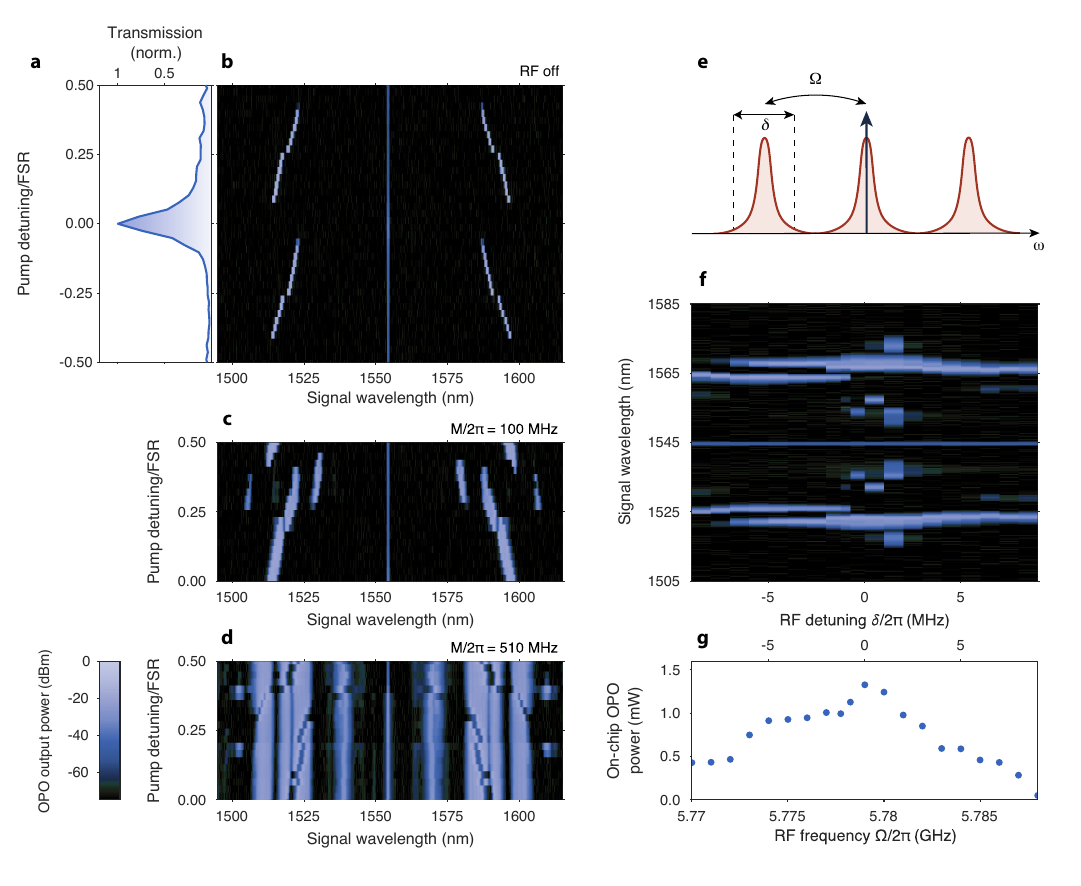}
  \end{center}
 \caption{\textbf{FM-OPO tuning with laser and RF detuning. }
\textbf{a},~Transmission of the leakage fundamental pump to the snail port of the resonator, corresponding to the OPO output in \textbf{b}.
\textbf{b},~Pump-wavelength tuning of the OPO in a nondegenerate regime, the output wavelength tuning curves repeat with a period of 1/2 FSR with respect to the pump wavelength.
\mbox{\textbf{c},~-\textbf{d}},~Pump-wavelength tuning of the FM-OPO driven with $M/2\pi\approx$ 100~MHz, and 510~MHz, respectively.
All measurements in \textbf{b}, \textbf{c}, and \textbf{d} correspond to pumping the device with about 140 mW of optical power. The faint line in the center of each colormap corresponds to the FH pump leakage into the cavity.
\textbf{e},~FM-OPO can be driven with the RF frequency on resonance with the cavity FSR near degeneracy ($\Omega = \zeta_1$) or detuned ($\Omega - \zeta_1 = \delta \neq 0$).
\textbf{f},~Tuning of the FM-OPO comb spectrum with RF detuning $\delta$.
\textbf{g},~Total integrated comb power as a function of RF frequency and detuning $\delta$.
}
 \label{fig:SI9}
\end{figure*}

\clearpage
\onecolumngrid

\section*{Supplementary Information}

\renewcommand{\figurename}{Supplementary Fig.}
\setcounter{figure}{0}

% \clearpage

\section{Optical parametric oscillator without modulation}

We model the doubly-resonant optical parametric oscillator (OPO) based on the Hamiltonian of the system. We separate it into the unperturbed and interaction part - $H_0$, $H_{\text{PA}}$:
\begin{eqnarray}
&&H_0 = \sum_n \omega(n) a_n^\ast a_n + \omega_bb^\ast b,
\label{eq:H0_1}
\\
&&H_\text{PA}=g \sum_n b a_n^\ast a_{-n}^\ast + \text{c.c.},
\label{eq:HPA_1}
\end{eqnarray}
where $a_n$ and $b$ correspond to the amplitudes of the $n$-th fundamental harmonic (FH) mode around the OPO degeneracy, point $n=0$, and the second harmonic (SH) pump mode. $\omega(n)$ and $\omega_b$ correspond to the frequency of the $n$-th FH mode and SH pump, $g$ is the $\chi^{(2)}$ nonlinear coupling rate. The coupled mode equations are given by:
\begin{eqnarray} 
\dot a_n &=& -\left(i\omega(n)+\cfrac{\kappa}{2} \right)a_n-2igba_{-n}^\ast
\label{eq:CME_1_a}
\\
\dot b &=& -\left(i\omega_b +\cfrac{\kappa_b}{2}\right)b-ig \sum_n a_n a_{-n}.
\label{eq:CME_1_b}
\end{eqnarray}
We can use this model to calculate the threshold and analyze the above-threshold behavior by assuming that $b$ is driven by a classical field at the pump frequency $\omega_p$. This leads to a drive term $H_{\text{d}} = \sqrt{\kappa_b} \beta_{in}e^{-i\omega_p t} b^\ast + \text{c.c.}$ For a doubly-resonant OPO, the loss of the $b$ field is dominated by the extrinsci coupler $\kappa_b^{(e)} \approx \kappa_b$. We remove the time dependence by putting $b$ in frame of $\omega_p$ (which assume is the same as $\omega_b$, and all $a_n$ in frame ${\omega_p}/{2}$, to maintain time-independence of $H_\text{PA}$). The resulting relevant parts of the Hamiltonian are:
\begin{eqnarray}
&&H_0 = \sum_n \left(\omega(n) - \cfrac{\omega_p}{2} \right) a_n^\ast a_n,
\label{eq:H0_2}
\\
&&H_\text{PA} = g \sum_n b a_n^\ast a_{-n}^\ast + \text{c.c.}
\label{eq:HPA_2}
\end{eqnarray}
and the coupled mode equations turn into:
\begin{eqnarray} 
\dot a_n &=& -\left(i\omega(n) - i \, \cfrac{\omega_p}{2} + \cfrac{\kappa}{2}
\right)
a_n-2igba_{-n}^\ast
\label{eq:CME_2_a}
\\
\dot b &=& -\cfrac{\kappa_b}{2} \, b - ig \sum_n a_n a_{-n}
+
i\sqrt{\kappa_b} \beta_{in}.
\label{eq:CME_2_b}
\end{eqnarray}

\subsection{Doubly-resonant OPO Threshold}

To find the threshold, we assume that the $a_n$’s are all equal to $0$, so $b$ obtains some complex field amplitude, and we obtain a system of two equations for each mode pair $(-n,+n)$. We also use the cavity dispersion:
\begin{eqnarray} 
\omega(n) = \omega_0+\zeta_1 n + \zeta_2 \cfrac{n^2}{2},
\label{eq:disp}
\end{eqnarray}
where $\zeta_1$, and $\zeta_2$ correspond to the first and second-order dispersion. This approach applied to equation \ref{eq:CME_2_a} yields:
\begin{eqnarray} 
\dot a_n &=& 
-\left[
i\left(
\omega_0+\zeta_1 n + \zeta_2 \cfrac{n^2}{2}
\right)
-i\cfrac{\omega_p}{2}+\cfrac{\kappa}{2}
\right]
a_n-2igba_{-n}^\ast,
\label{eq:CME_3_a}
\\
\dot a^\ast_{-n} &=& 
-\left[
-i\left(
\omega_0+\zeta_1 (-n) + \zeta_2 \cfrac{(-n)^2}{2} 
\right)
+i\cfrac{\omega_p}{2}+\cfrac{\kappa}{2}
\right]
a^\ast_{-n}+2igb^\ast a_{n}.
\label{eq:CME_3_a*}
\end{eqnarray}
From this system of equations, we have a coupled system involving two modes $a_n$ and $a_{-n}$. We can write the system of equations in a matrix form by treating $(a_n, a_{-n}^\ast)$ as a complex vector as $\text{d} \textbf{a}/\text{d}t = \mathbf{M} \mathbf{a}$, where $\mathbf{a} = \begin{pmatrix} a_n \\ a_{-n}^\ast \end{pmatrix}$ is the complex vector of amplitudes, and $\mathbf{M}$ is the matrix:
\begin{eqnarray} 
\mathbf{M} =
\begin{pmatrix}
-\left[
i \left( - \Delta +
\zeta_1 n + \zeta_2 \cfrac{n^2}{2}
\right)
+\cfrac{\kappa}{2}
\right] & -2igb \\
2igb^\ast & 
-\left[
-i\left( -\Delta
+\zeta_1 (-n) + \zeta_2 \cfrac{(-n)^2}{2} 
\right)
+\cfrac{\kappa}{2}
\right]
\end{pmatrix},
\label{eq:CME_matrix}
\end{eqnarray}
where we introduced the pump detuning defined as $\Delta = {\omega_p}/{2} - \omega_0$. This equation describes the evolution of the complex amplitudes $a_n$ and $a_{-n}^\ast$ in terms of a linear transformation defined by the matrix $\mathbf{M}$.

To find the stability conditions, one has to calculate the eigenvalues of the matrix $\mathbf{M}$ and find the conditions for the real parts of these eigenvalues to be negative. We can compute its eigenvalues by solving the characteristic equation $\text{det}(\mathbf{M} - \lambda \mathbf{I}) = 0$ that leads to the following quadratic equation:
\begin{eqnarray} 
\left[\lambda + i\left(-\Delta + \zeta_1 n + \zeta_2 \cfrac{n^2}{2}\right) + \cfrac{\kappa}{2} \right] \left[ \lambda - i\left( - \Delta +\zeta_1 (-n) + \zeta_2 \cfrac{(-n)^2}{2}\right)  + \cfrac{\kappa}{2} \right] - 4g^2|b|^2 = 0.
\label{eq:det-quadratic}
\end{eqnarray}
The corresponding stability criterion is:
\begin{eqnarray} 
16 g^2 |b|^2 > \kappa^2 + (2\Delta - n^2 \zeta_2)^2.
\label{eq:stability_crit}
\end{eqnarray}
The threshold of the OPO is minimized when the pump detuning perfectly compensates for the second-order dispersion for the $n$-th pair of modes $\Delta = n^2 \zeta_2/2$. In that case, we can substitute a steady-state solution for $b$ into the stability condition and see that the threshold of the doubly-resonant OPO is given by:
\begin{eqnarray}
P_{\text{th}} = \hbar\omega_{p} \, \cfrac{\kappa_a^2 \kappa_b}{64 g^2}.
\label{eq:P_th}
\end{eqnarray}

\subsection{Above-threshold behavior}

To find the relation between the pump power and the output power of the OPO we solve equations \ref{eq:CME_1_a}-\ref{eq:CME_1_b} in a steady state. Above the threshold, one pair of signal-idler modes will dominate the dynamics of the system, so we neglect the other FH modes:
\begin{eqnarray} 
&a_n = \cfrac{4igba_{-n}^\ast}{\kappa_a}
\label{eq:CME_ss_a}
\\
&b = \cfrac{-4ig a_n a_{-n}
+ 2i\sqrt{\kappa_b} \beta_{in}}{\kappa_b}
\label{eq:CME_ss_b}
\end{eqnarray}
Substituion of equation \ref{eq:CME_ss_b} into \ref{eq:CME_ss_a} yields:
\begin{eqnarray}
0 =  \cfrac{8g^2}{\kappa_b}\, a_n|a_{-n}|^2 
- \cfrac{4g}{\sqrt{\kappa_{b}}} \, \beta_{in} a_{-n}^{\ast}
+
\cfrac{\kappa_a}{2}\, a_n.
\label{eq:opo_quadratic_eq}
\end{eqnarray}
We can write an analogous equation for the signal and idler modes. Assuming that the amplitudes are real and the loss rates for the signal and idler modes are the same, we find the amplitudes of the signal and idler modes as
\begin{eqnarray}
|a_n|^2 = |a_{-n}|^2 =  \cfrac{\sqrt{\kappa_b}}{2g} \, \beta_{\text{in}}
-
\cfrac{\kappa_a \kappa_b}{16g^2}
\label{eq:opostability_crit}
\end{eqnarray}
The total output power of the doubly-resonant OPO is:
\begin{eqnarray}
P_{\text{out}} 
=
4 \eta_{a}
P_{\text{th}} \Bigg(
\sqrt{\cfrac{P_{\text{in}}}{P_{\text{th}}}} - 1
\Bigg),
\label{eq:P_out}
\end{eqnarray}
where $\eta_a = \kappa_a^{(e)}/\kappa_a$ is the cavity extraction efficiency for the FH modes and the input power is defined as \mbox{$P_{\text{in}} = \hbar \omega_p |\beta_{\text{in}}|^2$}. 

We can relate the OPO efficiency to the pump depletion by looking at the ${b}$ amplitude in a steady state (equation \ref{eq:CME_ss_b}). By substituting the solutions for the FH modes, we see that ${b} = i\kappa_a/4g$ and use the input-output relations to find the output amplitude:
\begin{eqnarray}
b_{\text{out}}
&=&
b_{\text{in}} + i\sqrt{\kappa_b^{(e)}} \, {b}
=
b_{\text{in}}
\Bigg(
1 - \cfrac{2 \sqrt{P_{\text{th}}}}{|b_\text{in}|}
\Bigg).
\label{eq:b_out}
\end{eqnarray}
The depletion of the pump power is:
\begin{eqnarray}
D 
=
4\,
\cfrac{P_{\text{th}}}{P_{\text{in}}}\, \Bigg(
\sqrt{\cfrac{P_{\text{in}}}{P_{\text{th}}}} - 1
\Bigg).
\label{eq:depl}
\end{eqnarray}
The OPO efficiency $\rho$ is proportional to depletion and the cavity extraction efficiency:
\begin{eqnarray}
\rho = \eta_a D.
\label{eq:efficiency_vs_depl}
\end{eqnarray}
We use this relationship to find the efficiency of our OPO and frequency comb generator. Note that the doubly-resonant OPO can achieve high efficiency by just increasing the coupling rate to the cavity and achieve $>50\%$ efficiency for $\kappa_a^{(e)}>\kappa_{a}^{(i)}$.

\subsection{Approximate single-mode model of a propagating pump field}

Our goal is to represent the propagating SH field and its dynamics approximately as the dynamics of a single $b$ mode. We note that this model can not capture complex spatial variations in the pump field. Let's first consider just the $b$ mode, ignoring the other $a_n$ modes of the system. We will need to define an  effective loss rate of $b$.  The input and output fluxes give the number of the photons within our waveguide in the steady state. After some time $T>\tau$, where $\tau$ is the amount of time it take the field to propagate across the waveguide, the number of photons in that region is given by
\begin{eqnarray}
|{b}(T>\tau)|^2
=
\int_{0}^{T} |\beta_{in}|^2\, dt
-
\int_{\tau}^{T } |\beta_{out}|^2\, dt,
\label{eq:b(t)}
\end{eqnarray}
where $\tau$ is the cavity round-trip time. If we neglect the propagation loss, we also see that $\beta_{in} = \beta_{out}$ and:
\begin{eqnarray}
|{b}(T>\tau)|^2
=
\int_{0}^{\tau} |\beta_{in}|^2\, dt
=
|\beta_{in}|^2 \, \tau.
\label{eq:b(t)2}
\end{eqnarray}
On the other hand, solving  equation \ref{eq:CME_1_b} in steady state and low-power approximation yields:
\begin{eqnarray}
b
=
- \cfrac{2}{\sqrt{ \kappa_{b}^{(e)} }} \, \beta_{in}.
\label{eq:b_ss_low_power}
\end{eqnarray}

We combine equations \ref{eq:b(t)}-\ref{eq:b_ss_low_power} to see that the effective loss rate of the $b$ mode is given by $\kappa_b = 4/\tau = 4 v_g / L$, where $v_g$ is the group velocity of the pump and $L$ is the total length over which the light propagates, here equivalent to the resonator length.

\clearpage

\section{Frequency-modulated optical parametric oscillator}

To include the effects of the intracavity phase modulator, we need to include additional term in the Hamiltonian:
\begin{eqnarray}
H_\text{mod}=2 M\cos (\Omega t)\sum_{\sigma=s,i}\sum_{m,m'} a^\ast_{\sigma,m'}a_{\sigma,m} + \text{c.c.},
\label{eq:H_mod_1}
\end{eqnarray}
where $M$ is the electro-optic modulation rate, and $\Omega$ corresponds to the RF frequency applied to the modulator. Applying RWA in the new frame, leads to:
\begin{eqnarray}
H_\text{mod}=M\sum_{\sigma=s,i}\sum_{m} \tilde a^\ast_{\sigma,m}\tilde a_{\sigma,m+1} + \text{c.c.}.
\label{eq:H_mod_2}
\end{eqnarray}
Resulting coupled mode equations, including both the parametric gain and the modulation, are then:
\begin{eqnarray}
H = H_0 + H_\text{PA} + H_\text{mod}.
\label{eq:CME_full_1}
\end{eqnarray}
The coupled mode equations in the text are generated from this classical Hamiltonian.

\subsection{Relabeling the modes according to offset from NDOPO signal and idler}

Once we have a nondegenerate oscillating solution for the equations above, the signal and idler oscillations would be at a specific value of $\pm n_\text{osc}$. We will also denote the frequencies of these oscillations as 
$\omega_\text{s} = \omega(+n_\text{osc})$ and
$\omega_\text{i} = \omega(-n_\text{osc})$.
The oscillating modes are then 
\begin{eqnarray} 
a_{\text{s},0} &\equiv& a_{+n_\text{osc}}\\
a_{\text{i},0} &\equiv& a_{-n_\text{osc}}.
\end{eqnarray}

We can count out from these oscillating modes with a new index variable, $m$:
\begin{eqnarray} 
a_{\text{s},m} &\equiv& a_{n_\text{osc}+m},\\
a_{\text{i},m} &\equiv& a_{-(n_\text{osc}+m)},
\end{eqnarray}
with frequencies
\begin{eqnarray}
\omega_{\text{s}}(m) &=& \omega(n_{\text{osc}}+m),\\
\omega_{\text{i}}(m) &=& \omega(-n_{\text{osc}}-m).
\end{eqnarray}

These can be written more explicitly using the original definition of $\omega(n)$:
\begin{eqnarray}
\omega_{\text{s}}(m) &=& \omega_0+\zeta_1 (n_{\text{osc}}+m) + \frac{\zeta_2}{2} (n_{\text{osc}}+m)^2,\\
\omega_{\text{i}}(m) &=& \omega_0+\zeta_1 (-n_{\text{osc}}-m) + \frac{\zeta_2}{2} (-n_{\text{osc}}-m)^2.
\end{eqnarray}

Now we rewrite Hamiltonians in terms of these new frequencies and the newly defined modes $a_{\text{s},m}$ and $a_{\text{i},m}$. For example, the zeroth order Hamiltonian $H_0$ would be:
\begin{equation}
H_0 = \sum_m \left[(\omega_{\text{s}}(m)-\omega_p/2) a_{\text{s},m}^\ast a_{\text{s},m} + (\omega_{\text{i}}(m)-\omega_p/2) a_{\text{i},m}^\ast a_{\text{i},m}\right].
\end{equation}
Similarly, the parametric amplification Hamiltonian $H_{\text{PA}}$ becomes
\begin{equation}
H_{\text{PA}} = g \sum_m (b a_{\text{s},m}^\ast a_{\text{i},m}^\ast + \text{c.c.})
\end{equation}
This relabeling puts the oscillating modes at the center  $m=0$ and allows us to examine the dynamics in their vicinity. The indices $m$ are the offsets from these central modes.

\subsection{Calculating a recurrence relation for the modal amplitudes}

First, we consider the steady state equation for the $b$ mode as:
\begin{equation}
\dot{b} = -\frac{\kappa_b}{2} b -i g\sum_m \tilde{a}_{\text{s},m}\tilde{a}_{\text{i},m}+i\sqrt{\kappa_b} b_{\text{in}}.
\end{equation}
At steady state, we expect the amplitude of the intracavity pump field to become:
\begin{equation}
b = -\frac{{2}i g}{\kappa_b}\sum_m \tilde{a}_{\text{s},m}\tilde{a}_{\text{i},m}+\frac{2i}{\sqrt{\kappa_b}} b_{\text{in}}
\end{equation}

For the idler and signal modes, we then derive the equations:
\begin{eqnarray}
\frac{d\tilde{a}_{\text{i},m}}{dt} &=& -\frac{\kappa_a}{2}\tilde{a}_{\text{i},m}-i \left((n_{\text{osc}}\zeta_2 m + m^2\zeta_2/2 )\tilde{a}_{\text{i},m} + g b \tilde{a}_{\text{s},m}^\ast + M(\tilde{a}_{\text{i},m+1}+\tilde{a}_{\text{i},m-1}) \right), \\
\frac{d\tilde{a}_{\text{s},m}}{dt} &=& -\frac{\kappa_a}{2}\tilde{a}_{\text{s},m}-i \left((n_{\text{osc}}\zeta_2 m + m^2\zeta_2/2 )\tilde{a}_{\text{s},m} + g b \tilde{a}_{\text{i},m}^\ast + M(\tilde{a}_{\text{s},m+1}+\tilde{a}_{\text{s},m-1}) \right).
\end{eqnarray}

Now we can find relations for steady-state amplitudes (assuming we can choose the phases so that all the amplitudes are real) -- the imaginary part of the equations above are given by:
\begin{eqnarray}
0 &=& (n_{\text{osc}}\zeta_2 m + m^2\zeta_2/2 )\tilde{a}_{\text{i},m}  + M(\tilde{a}_{\text{i},m+1}+\tilde{a}_{\text{i},m-1}),\label{eqn:reccurence_1} \\
0 &=& (n_{\text{osc}}\zeta_2 m + m^2\zeta_2/2 )\tilde{a}_{\text{s},m}  + M(\tilde{a}_{\text{s},m+1}+\tilde{a}_{\text{s},m-1})\label{eqn:reccurence_2} ,
\end{eqnarray}

Notice that we used $\text{Re}[g b \tilde{a}_{\text{i},m}^\ast] = 0$, i.e.,  if $b_{\text{in}}$ is chosen to be real, $b$ will be imaginary.

\subsection{Verifying that the FM solution satisfies the equations of motion}

We show that a solution of the form
\begin{equation}
\tilde{a}_{\text{i},m}=A_{\text{i}} J_m(\Gamma),~~~~\tilde{a}_{\text{s},m}=A_{\text{s}} J_m(\Gamma)
\end{equation}
approximately satisfies the steady-state dynamics of the system derived above. Substituting the relations into the equations~(\ref{eqn:reccurence_1}) and (\ref{eqn:reccurence_2}), 
we find:
\begin{equation}
0 = - (n_{\text{osc}}\zeta_2 m + m^2\zeta_2/2 )A_i J_m(\Gamma) +M(A_i J_{m+1}(\Gamma)+A_i J_{m-1}(\Gamma)).
\end{equation}

We now use the Bessel function recurrence relation to simplify this expression. The recurrence relation for Bessel functions is:
\begin{equation}
\frac{2m}{x}J_{m}(x) = J_{m-1}(x) + J_{m+1}(x),
\end{equation}
which leads to
\begin{equation}
0 = - (n_{\text{osc}}\zeta_2 m + m^2\zeta_2/2 ) +M \cdot \frac{2m}{\Gamma}.
\end{equation}
Obviously, it is not possible to satisfy this relation exactly due to the dependence on $m$ of inside equation. But, we can make this \emph{approximately} true by setting
\begin{equation}
\Gamma = \frac{2 M}{n_{\text{osc}}\zeta_2}.
\end{equation}
The bandwidth is then given by
\begin{equation}
\text{B.W.}\equiv2\Gamma\Omega = \frac{4M\Omega}{n_{\text{osc}}\zeta_2}
\end{equation}

We have solved the comb dynamics in frequency domain. To find the time-domain solution, we use the Jacobi-Anger relation
\begin{equation}
e^{iz\sin\theta} = \sum_{m=-\infty}^{\infty} J_m(z) e^{im\theta}
\end{equation}
we find that the steady-state solution of the system appears as swept signal and idler tones. These would be represented in the form
\begin{eqnarray}
a_\text{i}(t) &=& A_\text{i}e^{-i\omega_\text{i} t}e^{i\Gamma \sin(\Omega t)} e^{i\omega_\text{p}t/2}\\
a_\text{s}(t) &=& A_\text{s}e^{-i\omega_\text{s} t}e^{-i\Gamma \sin(\Omega t)}e^{i\omega_\text{p}t/2}.
\end{eqnarray}

\section{Numerical modeling}

\subsection{Quasi-static approximation}
We solve the coupled mode equations for 2,500 modes. To increase the numerical efficiency of the ODE solver, we notice that the extrinsic coupling rate of the ${b}$ amplitude is much faster than any other rates in the system and introduce a quasi-static approximation. During each step of the ODE solver, we assume that the SH mode is in a steady state, which yields:
\begin{eqnarray}
\dot{{a}}_n
&=&
\Bigg[
i\Bigg(
\Delta + n\delta - \cfrac{n^2\zeta_2}{2}
\Bigg)
-\cfrac{\kappa_{a,n}}{2}\, 
\Bigg]
{a}_{n}
-
i{M}
\Big(
{a}_{n-1} 
+
{a}_{n+1} 
\Big)
- 2ig
{a}_{-n}^{\ast} {b} 
\\
{{b}} 
&=&
- 2i \cfrac{g \sum_{n} {a}_{n} {a}_{-n}
-
i\sqrt{\kappa^{(e)}_b} \beta_{in}}{{\kappa_b^{(e)}}}.
\label{eq:quasi-static}
\end{eqnarray}
These are the equations we solve in the main text to predict the shape of the optical frequency combs generated in our device and their total comb count.

\end{document}